\documentclass[conference]{IEEEtran}
\IEEEoverridecommandlockouts
\usepackage{cite}
\usepackage{amsmath,amssymb,amsfonts}
\usepackage{algorithmic}
\usepackage{graphicx}
\usepackage{textcomp}
\usepackage{xcolor}
\usepackage{subfigure}
\usepackage[ruled,linesnumbered]{algorithm2e}
\usepackage{booktabs} 
\usepackage{multirow} 
\usepackage{tcolorbox}
\usepackage{colortbl}
\usepackage{url}
\usepackage{hyperref}
\hypersetup{
hidelinks
}

\def\BibTeX{{\rm B\kern-.05em{\sc i\kern-.025em b}\kern-.08em
    T\kern-.1667em\lower.7ex\hbox{E}\kern-.125emX}}
\begin{document}

\title{\emph{DeepScaler}: Holistic Autoscaling for Microservices Based on Spatiotemporal GNN with Adaptive Graph Learning}

\makeatletter
\newcommand{\ssymbol}[1]{^{\@fnsymbol{#1}}}
\makeatother

\author{
\IEEEauthorblockN{Chunyang Meng$\ssymbol{2}$, Shijie Song$\ssymbol{2}$, Haogang Tong$\ssymbol{2}$, Maolin Pan$\ssymbol{3}$ and Yang Yu{$\ssymbol{3}$$^{,}$$\ssymbol{1}$}}
\IEEEauthorblockA{$\ssymbol{2}$\textit{School of Computer Science and Engineering, Sun Yat-sen University, Guangzhou, China} \\
 $\ssymbol{3}$\textit{School of Software Engineering, Sun Yat-sen University, Zhuhai, China}\\
$\ssymbol{1}$\textit{Author to whom correspondence should be addressed} \\
Email: \{mengchy3, songshj6, tonghg\}@mail2.sysu.edu.cn, \{panml, yuy\}@mail.sysu.edu.cn} 
}

\maketitle

\begin{abstract}
Autoscaling functions provide the foundation for achieving elasticity in the modern cloud computing paradigm. It enables dynamic provisioning or de-provisioning resources for cloud software services and applications without human intervention to adapt to workload fluctuations. 
However, autoscaling microservice is challenging due to various factors. 
In particular, complex, time-varying service dependencies are difficult to quantify accurately and can lead to cascading effects when allocating resources.
This paper presents \emph{DeepScaler}, a deep learning-based holistic autoscaling approach for microservices that focus on coping with service dependencies to optimize service-level agreements (SLA) assurance and cost efficiency. 
\emph{DeepScaler} employs (i) an expectation-maximization-based learning method to adaptively generate affinity matrices revealing service dependencies and (ii) an attention-based graph convolutional network to extract spatio-temporal features of microservices by aggregating neighbors' information of graph-structural data. Thus \emph{DeepScaler} can capture more potential service dependencies and accurately estimate the resource requirements of all services under dynamic workloads. It allows \emph{DeepScaler} to reconfigure the resources of the interacting services simultaneously in one resource provisioning operation, avoiding the cascading effect caused by service dependencies.
Experimental results demonstrate that our method implements a more effective autoscaling mechanism for microservice that not only allocates resources accurately but also adapts to dependencies changes, significantly reducing SLA violations by an average of 41\% at lower costs.

\end{abstract}


\section{Introduction}
\label{sec_introduction}

Cloud computing offers a well-consolidated paradigm for on-demand services provisioning on a pay-as-you-go basis~\cite{coutinho2015elasticity}. It has gained more popularity in the last decade, as evidenced by the widespread use of cloud-based software services and applications~\cite{chen2016self}.
One of its central attributes is elasticity, which enables flexible provisioning or de-provisioning computing resources according to customers' needs~\cite{al2017elasticity}.
To efficiently employ the elasticity, acquiring and releasing cloud resources automatically and timely is vital, as human intervention is difficult or even impossible~\cite{chen2018survey}. This mechanism of automatic and rapid scale outward and inward commensurate with demands under fluctuating workloads to guarantee quality of service (QoS) at minimal costs is called autoscaling~\cite{qu2018auto}.

\begin{figure}[htbp]
    \includegraphics[width = 1\linewidth]{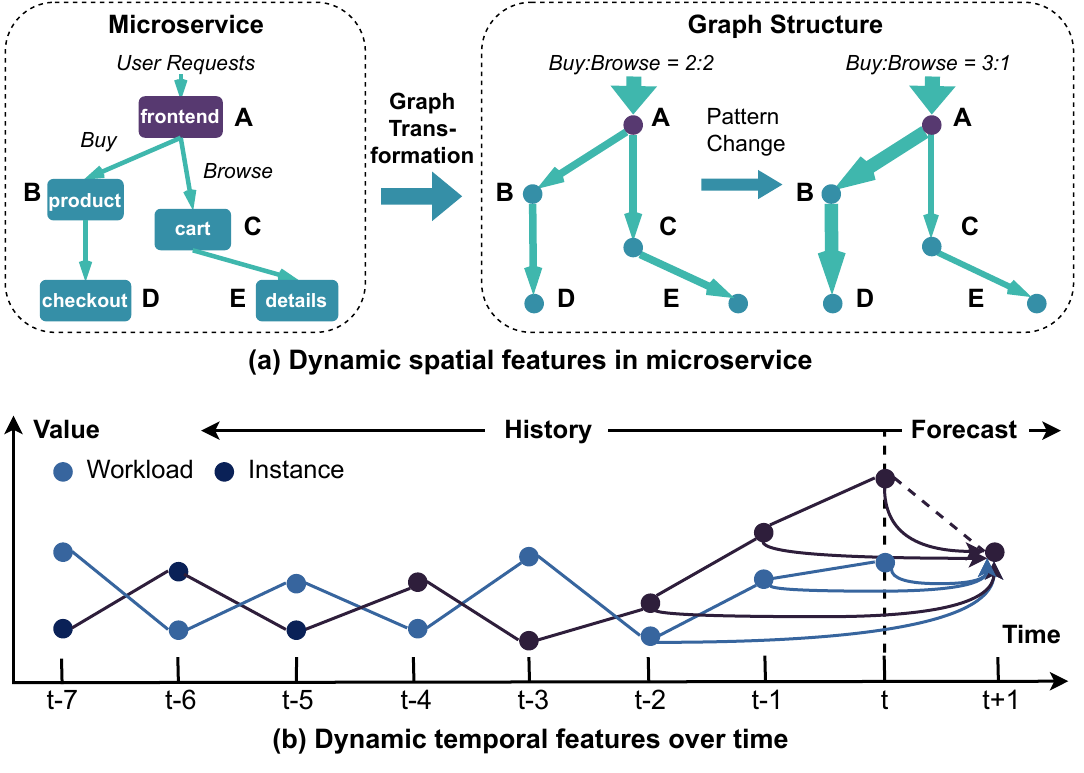}
    \caption{Dynamic Spatial-temporal Features in Microservice
        }
    \label{fig_feature}
\end{figure}

As cloud applications continue to grow in size and complexity, traditional monolithic applications face the problem of high development and maintenance costs~\cite{zhong2022machine}. The microservice architecture is proposed to address this problem, which splits single-component applications into multiple loosely coupled and self-contained microservice components~\cite{dragoni2017microservices}. Nonetheless, designing and implementing an efficient autoscaler for microservices is a challenging task due to various factors. Specifically, the primary challenges are as follows:
\begin{itemize}
     \item \textbf{Complex dependencies.} In microservice applications, there are complex dependencies between internal services that can cause cascading effects when provisioning resources for interrelated services~\cite{yang2019miras}. 
    \item \textbf{Dynamic environments.} Cloud-native microservices serve dynamic workloads with time-varying patterns and frequent software updates of microservices lead to changes of inter-dependencies~\cite{gan2019open}.
     \item \textbf{QoS and cost trade-off.} There is a trade-off between QoS and costs, since over-provisioning leads to increased costs, while under-provisioning violates service-level agreements (SLA).
     ~\cite{singh2019research}. 
\end{itemize}

\textbf{Research Gaps.} 
Extensive related works have been proposed on autoscaling for microservices, including rule-based~\cite{DARADKEH2023102713,mirhosseini2021parslo,horovitz2018efficient,urgaonkar2008agile} and learning-based~\cite{zhang2020sarsa,zafeiropoulos2022reinforcement,cai2023automan} methods.
However, these works mainly confront two problems. 
First, most of them are insensitive to graph information that cannot effectively capture spatio-temporal characteristics of microservices (Fig.~\ref{fig_feature}), usually scaling resources based on performance bottleneck services independently instead of holistically, resulting in distorted modeling and suboptimal resource allocation.
Secondly, partial approaches consider service dependencies only in surface invocation relationships or rely on prior-knowledge-based handcraft dependency graphs. They are usually inadequate and lack adaptive extraction of other latent dependencies (e.g., causal relationships) hidden in the real data.

\textbf{The \emph{DeepScaler}.} To fill these gaps and address the challenges, we present a deep learning-based holistic autoscaling method for microservice, namely \emph{DeepScaler}. It aims to automatically re-provision resources of all scaling-needed services simultaneously according to precise resource estimation on dynamic workloads to optimize runtime QoS assurance and cost efficiency. 
\emph{DeepScaler} consists of four procedures, including a performance monitor, a resource estimator, an adaptive learning module, and an online scheduler.
\emph{DeepScaler} first uses the performance monitor to collect and store tracing and telemetry data of microservices. Then, it uses the adaptive learning module to train the resource estimator and learn affinity matrixes that reveal service dependencies. Finally, it periodically uses the trained resource estimator, the learned affinity matrix, and the collected real-time performance indicators to predict future resource demands and allocate resources accordingly.

\textbf{Contributions.} 
To the best of our knowledge, this is the first work to provide holistic autoscaling for large-scale microservice applications by using spatio-temporal neural network with adaptive graph learning. 
Generally, our main contributions are: 
\begin{itemize}
    \item \textbf{\emph{Resource Estimation}:} We present a deep learning-based resource estimation method for microservices. It utilizes attention based graph neural network that can capture the spatio-temporal features of microservices to estimate resource, making the estimation more accurate.
    \item \textbf{\emph{Adaptive Learning}:} We propose an adaptive learning method based on expectation maximization (EM) and parametric graph learning for the above estimation model, which can capture latent service dependencies adaptively.
    \item \textbf{\emph{Holistic Autoscaling}:} We present \emph{DeepScaler}, holistic autoscaling for microservices, simultaneously allocating resources for interacting services in one scaling action, mitigating SLA violations while costing less.
    \item \textbf{\emph{Implementation \& Evaluation}:} We implement and evaluate \emph{DeepScaler} based on an Istio-enabled Kubernetes container orchestration system in a public cloud environment. The experimental result shows that \emph{DeepScaler} is a promising approach.
\end{itemize}

\section{Background \& Characterization}
\label{sec_background}
\newtheorem{definition}{\hspace{0em}\textbf{\emph{Definition}}}[section]

\subsection{Microservice}
\begin{figure}[htbp]
    \centerline{\includegraphics[width = 0.85\linewidth]{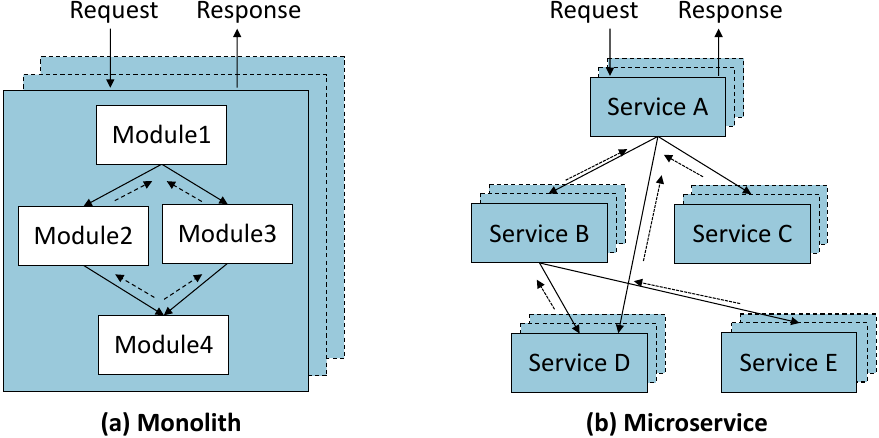}}
    \caption{Monolith and Microservice Architectures}
    \label{fig_microservice}
\end{figure}

Currently, cloud-based software services and applications have acquired popularity, encompassing all spheres of human activity. Many of these applications require interactivity and have strict performance requirements (high throughput and minimal tail latency) while also needing to ensure availability and manage frequent software updates~\cite{delimitrou2016hcloud,delimitrou2015tarcil}.
Large cloud providers such as Amazon, Twitter, Netflix, Apple, and eBay have adopted a significant design change to address the challenge of managing the constantly increasing size and complexity of their systems and meeting conflicting demands by moving away from traditional monolithic architecture and instead using graphs consisting of tens or hundreds of single-purpose, loosely-coupled microservices~\cite{microservicesWorkshop,evolutionMicroservices}. Fig.~\ref{fig_microservice} illustrates the difference between monolithic and microservices architectures.

Several compelling reasons substantiate the surging popularity of microservices. First, they foster composable software design, streamlining and accelerating development processes, with each microservice catering to a small subset of the application's functionality. Second, microservices facilitate programming language and framework heterogeneity, allowing each tier to develop in the most suitable language, with only a standard application programming interface (API), e.g., remote procedure calls (RPC)~\cite{grpc} or representational state transfer (RESTful)~\cite{richardson2008restful}, required for inter-microservice communication.
Finally, microservices simplify the resolution of correctness and performance issues, as bugs can isolate to specific tiers, unlike monoliths, where rectifying bugs necessitates troubleshooting the entire service.
Despite the numerous benefits of microservices, they substantially challenge many assumptions on the design of current cloud systems. As a result, they pose both opportunities and challenges in terms of optimizing QoS and utilization.

\subsection{Graph Neural Network}
Graph Neural Network (GNN) is a type of neural networks that learn and infer information from graph-structured data~\cite{zhou2020graph}. It combines graph embedding techniques and neural network operators to learn latent representations for nodes in a graph by taking into account the features of their neighboring node~\cite{wu2020comprehensive}. GNNs have gained significant attention in recent years due to its success in a variety of tasks such as node classification, link prediction, and social network analysis, among others. 

In GNNs, input involves a graph represented by an adjacency matrix and a feature matrix.
The adjacency matrix captures the relationships between nodes, while the feature matrix encapsulates the attributes or characteristics of each node.
For example, in a social graph, nodes represent users with feature vectors denoting characteristics like age, location, and preferences, and links between users signify relationships like friendships. 
The output of a GNN model is termed the representation vector or embedding, a multi-dimensional vector encapsulating latent insights about nodes. 
For the social graph just mentioned, a user's embedding could be a numeric vector reflecting their attributes. 
By transforming the intricate semantics of raw data into well-structured numerical forms, these embeddings facilitate an array of downstream analytical tasks, including user identity classification~\cite{zhang2022deep}, friendship prediction~\cite{sankar2021graph}, and community detection~\cite{chen2017supervised}.

\begin{figure}[htbp]
    \centerline{\includegraphics[width = 1\linewidth]{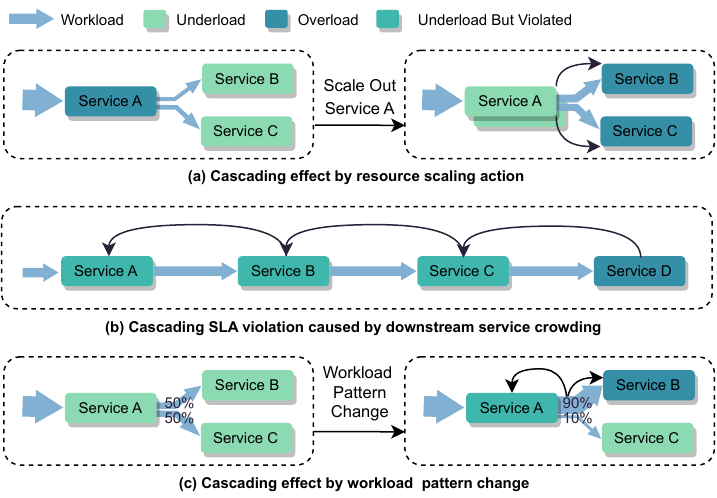}}
    \caption{Cascading Effects in Microservices}
    \label{fig_interaction}
\end{figure}

\subsection{Key Insights}\label{sec_motivation}
As mentioned earlier in Research Gaps of section \ref{sec_introduction}, although there have been some studies on microservice autoscaling, they lack more careful considerations on service dependencies, limiting the performance of their algorithms. 
To better understand the dependencies between microservices and study SLA violation characteristics, we have run extensive autoscaling experiments on widely used microservice benchmarks (i.e. Train-Ticket~\cite{trainticket}, Online-Boutique~\cite{boutique} and BookInfo~\cite{bookinfo}) in a public cloud environment. Our key insights are as follows.

\textbf{Insight 1: Holistic Autoscaling is Necessary.}
Properly selecting the exemplary service for resource scaling in microservices is critical to mitigating SLA violations. Some algorithms~\cite{yu2019microscaler,horovitz2018efficient,jiang2010autonomous} choose to scale the current performance bottleneck service, but our study shows that scaling only the bottleneck service may take more time to mitigate SLA violations. 
As shown in Fig.~\ref{fig_interaction}(a), consider that service A depends on its downstream services B and C. When resources are allocated to bottleneck service A with a large workload, A's ability to process requests increases, and the request rate at which A dispatched downstream by A increases, leading to the possibility that downstream services B and C may become new bottleneck services, resulting in an inability to mitigate SLA violations immediately.

\begin{center}
    \begin{tcolorbox}[colback=gray!10,
                      colframe=black,
                      width=\linewidth,
                      arc=2mm, auto outer arc,
                      boxrule=1pt,
                     ]
        \begin{definition}
            \label{def_holistic}
            \emph{Holistic autoscaling} is a kind of autoscaling method that allows inter-related services to simultaneously re-provision during resource allocation to reduce the cascading impact caused by service dependencies. 
        \end{definition}
    \end{tcolorbox}
\end{center}

Although the SLA violations can be completely mitigated by multiple scaling actions to eliminate the bottleneck services, it undoubtedly prolongs the SLA violation time. Our study indicates that, in microservice autoscaling, the optimal approach should consider service dependencies and interrelated microservices should be scaled simultaneously to avoid bottleneck shifting, thus mitigating SLA violations immediately.

\textbf{Insight 2: Service dependencies must be discovered adaptively.} 
Recent approaches have explored service dependencies of microservices and established corresponding holistic autoscaling methods to guarantee runtime QoS and reduce resource consumption. For example, Meng et al.~\cite{meng2022hra} automatically captures potential service dependencies from global application-level metrics via deep reinforcement learning (DRL) to guide scaling decisions, Tong et al.~\cite{tong2021holistic} constructs balanced queuing networks based on expert-driven predefined dependency graphs that enable scaling all services simultaneously, and Yu et al.~\cite{yu2020microscaler} automatically constructs service invocation relationships via online telemetry traces to determine which services need to be scaled.

\begin{center}
    \begin{tcolorbox}[colback=gray!10,
                      colframe=black,
                      width=\linewidth,
                      arc=2mm, auto outer arc,
                      boxrule=1pt,
                     ]
        \begin{definition}
        \label{def_dependency}
        \emph{Latent service dependencies} belong to service dependencies, which describe the dependencies adaptively discovered from actual data rather than provided as ground truth knowledge. They may be some highly abstract dependencies that are difficult to understand.
        \end{definition}
    \end{tcolorbox}
\end{center}

However, our study shows that this is not sufficient by itself. The requirement is to capture the time-varying dependencies adaptively and as accurately as possible. Traditional neural networks are insensitive to graph structure information leading to modeling distortions. Expert-driven approaches face expensive model reconstruction engineering and difficulty adapting to dependency changes. 
The automatically constructed invocation graph based on telemetry traces only captures a type of communication-based dependencies between microservice instances, which cannot adequately describe other \emph{latent service dependencies} (e.g., business correlations).

\textbf{Insight 3: A single indicator, including request workload, resource utilization, and end-to-end latency, cannot identify the need for scaling.} 
Many microservice autoscaling algorithms~\cite{Quality} determine the need for scaling based on a single performance indicator (e.g., workload, resource utilization, and end-to-end latency). However, our study shows that a single performance indicator is insufficient to identify the need for scaling. We explain below the reasons why using each of these three types of metrics alone is not sufficient.

First, it is not sufficient to use the size of resource utilization (e.g., CPU) to trigger scaling actions. For example, when a service has a high CPU utilization, it does not mean it needs to scale out. A software bug, such as an infinite loop, could cause it. Conversely, when the CPU utilization of a service is low, it does not mean that the service needs to scale in either because an unacceptable response latency may be caused by the limited availability of other underlying computational resources (e.g., memory) that need to be scaled up.

Second, as the end-to-end latency of a microservice depends not only on its own response time but also on the response time of its downstream services, it leads to confusion in deciding whether scaling is required based on the end-to-end latency. As shown in Fig.~\ref{fig_interaction}(b), when a downstream service is crowded, cascading SLA violations occur for the upstream services in its invocation chain.

Finally, since microservice workloads have dynamic load patterns, workloads will be dispatched to different downstream services in varying manner, making it impractical to determine the need for scaling based on the intensity of the workload. Consider service A and a large workload, where service A depends on downstream services B and C. 
As shown on the left of Fig.~\ref{fig_interaction}(c), the large workload is evenly distributed to different downstream services, resulting in service A providing an acceptable quality of service.
Conversely, as shown on the right of Fig.~\ref{fig_interaction}(c), when the workload is congested on the same downstream service B, it may lead to an SLA violation in the response time of service A.

\section{Problem Formulation And System Design}

\subsection{Problem Formulation}\label{sec_problem_formulation}
In this study, a microservice application is considered as a graph containing multiple interrelated services,
denoted by $G=(V,A)$, where $V$ is a finite set of nodes representing the services with $|V|=N$, and $A \in \mathbb{R}^{N \times N}$ represents the adjacency matrix of the service dependency. Each node in the microservice application $G$ produces a feature vector of length $C$ at each time step, where $C$ is the number of online telemetry performance indicators collected with the same sampling frequency $F$.

\begin{figure}[htbp]
    \centerline{\includegraphics[width = 0.92\linewidth]{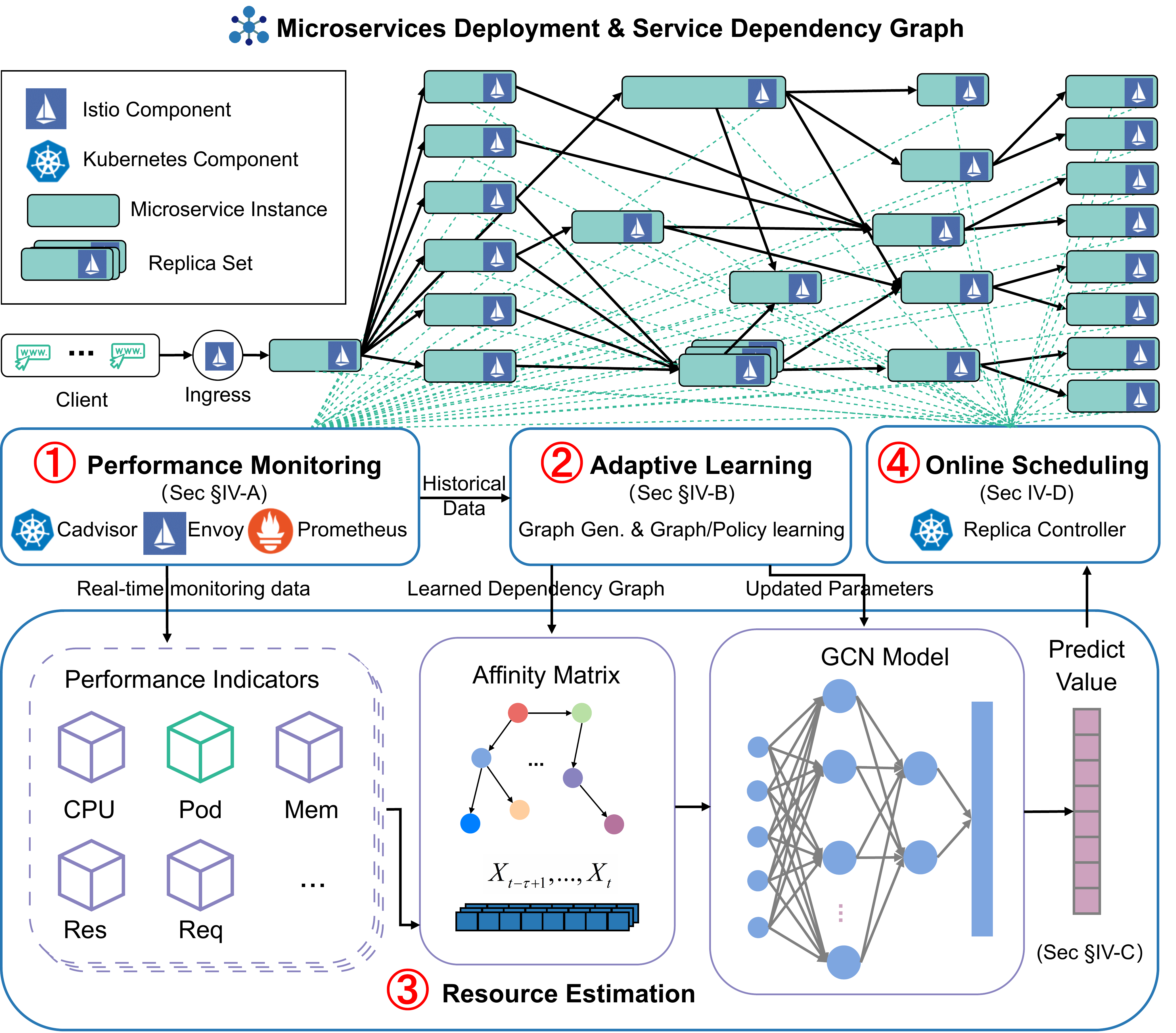}}
    \caption{System Overview}
    \label{fig_overview}
\end{figure}

\textbf{The Problem of Microservice Autoscaling.}
Assume the $c$-th time series recorded on each node in the microservice application $G$ is the instance replicas sequence, and $c \in (1, \ldots, C)$. We use $\mathbf{x}_{t}^{k,i}\in \mathbb{R}^{C}$ denotes the values of the $k$-th feature of node $i$ at time $t$, and $\mathbf{x}_{t}^{i}\in \mathbb{R}^{N\times C}$ denotes the values of all the features of node $i$ at time $t$. 
$\mathbf{X}_{t}=\left(\mathbf{x}_{t}^{1}, \mathbf{x}_{t}^{2}, \ldots, \mathbf{x}_{t}^{N}\right)^{T} \in \mathbb{R}^{N \times C}$ denotes the values of all the features of all the nodes at time $t$. 
$\mathcal{X}_t=\left(\mathbf{X}_{t-\tau+1}, \ldots, \mathbf{X}_{t-1}, \mathbf{X}_{t}\right)^{T} \in$ $\mathbb{R}^{N \times C \times \tau}$ denotes the values of all the features of all the nodes on the past $\tau$ time slices at time $t$.
At each discrete time step
$t=0,1,2\ldots$,T, given $\mathcal{X}_t$, the goal is to find a sequence of instance replicas $\mathbf{Y}_t=\left(\mathbf{y}_t^{1}, \mathbf{y}_t^{2}, \ldots, \mathbf{y}_t^{N}\right)^{T} \in \mathbb{R}^{N }$ of all the nodes on the whole microservice application to minimize the SLA violation rate $R$ with less resource cost $Cost$ in the total executing time, where $\mathbf{y}_t^{i}=x_{t+1}^{c,i} \in \mathbb{R}$
denotes the required instance replicas of node $i$ at time $t+1$. The definitions of $R$ and $Cost$ are shown in \eqref{eq_slavio} and \eqref{eq_cost}, respectively.
\begin{equation}\label{eq_slavio}
    R = T_{violation}/T  
 \end{equation}
 \begin{equation}\label{eq_cost}
     Cost = \frac{F}{T^2}\sum_{t=0}^{T}\sum_{i=1}^{N}c_t^i\times y_t^i   
 \end{equation}
where $T_{violation}$ is the cumulative time that the end-to-end response time of the microservice application violates the SLA requirements.
$c_t^i$ is the number of CPU logical cores configured for each instance of service $i$ at time $t$.

\subsection{System Design}

To address the above problems, we propose a novel holistic autoscaling method based on spatio-temporal GNN and adaptive learning, namely \emph{DeepScaler}, with the overview shown in Fig.~\ref{fig_overview} and pseudocode shown in Algorithm~\ref{alg_autoscaling}, which achieves runtime SLA guarantee and cost efficiency for microservices in dynamic environments. 
The system of \emph{DeepScaler} is designed as a tight MAPE (Monitoring, Analysis, Planning, and Execution) control loop.
We briefly introduce each phase here; the details can be found in §\ref{sec_method}.

\begin{algorithm}
    \caption{Pseudocode of \emph{DeepScaler}}
    \label{alg_autoscaling}
    \begin{algorithmic}[1]
        \STATE Initialize GNN-based resource estimator $P$ with parameters $\theta$ and adjacency matrix $A$
        \STATE Initialize telemetry dataset $\mathcal{D}$ and learning interval $C$ 
        \WHILE{not done}
        \STATE Observe current states $\mathbf{X}_{t}=\left(\mathbf{x}_{t}^{1}, \mathbf{x}_{t}^{2}, \ldots, \mathbf{x}_{t}^{N}\right)^{T}$
        and store $\mathbf{X}_t$ in $\mathcal{D}$ according to §\ref{sec_method_performance_monitoring}
        \STATE Every $C$ time steps perform adptive learning (§\ref{sec_method_adaptive_learning}):\\ \ \ \ \ \ \ \ \ \ \ 
        $A,P_{\theta}=AdaptLearning(A,P_{\theta};\mathcal{D})$
        \STATE Query last $\tau$ observations: \\ \ \ \ \ \ \ \ \ \ \ \ \ 
        $\mathcal{X}_t=\left(\mathbf{X}_{t-\tau+1}, \ldots, \mathbf{X}_{t-1}, \mathbf{X}_{t}\right)^{T}$ 
        \STATE Predict future resource needs (§\ref{sec_method_resource_estimation}): \\
        \ \ \ \ \ \ \ \ \ \ \ \ \ \ \ \ \ \ $\mathbf{Y}_t=P_{\theta}(\mathcal{X}_t,A)$
        \STATE Allocate resources according to $\mathbf{Y}_t$ (§\ref{sec_method_online_scheduling})
        \ENDWHILE
    \end{algorithmic}  
\end{algorithm}

\begin{enumerate}

    \item \textbf{Monitoring.} 
    \emph{DeepScaler} first needs to monitor some performance indicators to determine whether scaling operations are necessary and how they should be performed. It does so using the \emph{Performance Monitor}, which is marked as \textcolor{red}{\textbf{\large\textcircled{\small 1}}} and described in §\ref{sec_method_performance_monitoring}. The \emph{Performance Monitor} collects tracing and telemetry data from every service instance in the microservice application and stores them in a centralized time-series database for processing. 

    \item \textbf{Analysis.} 
    Then, the collected data are further processed in the analysis phase. \emph{DeepScaler} periodically uses the \emph{Adaptive Learning Module} (marked as \textcolor{red}{\textbf{\large\textcircled{\small 2}}} in Fig.~\ref{fig_overview} and described in §\ref{sec_method_adaptive_learning}) with all the recently collected tracing and telemetry data to learn affinity matrices revealing the time-varying service dependencies and train the GNN-based \emph{Resource Estimator}.

    \item \textbf{Planning.}
    The planning phase estimates the resource demands of coming workloads for SLA assurance and resource efficiency optimization.
    By using the real-time data collected in \textcolor{red}{\textbf{\large\textcircled{\small 1}}} and the dependency graph learned in \textcolor{red}{\textbf{\large\textcircled{\small 2}}}, and the fine-trained \emph{Resource Estimator} (marked as \textcolor{red}{\textbf{\large\textcircled{\small 3}}} and described in §\ref{sec_method_resource_estimation}), \emph{DeepScaler} makes online predictions of the instances required at the next time step for every service in the microservice.
    
    \item \textbf{Execution.} 
    Finally, in the execution phase,
    \emph{DeepScaler} utilize the \emph{Online Scheduler} (marked as \textcolor{red}{\textbf{\large\textcircled{\small 4}}} and described in §\ref{sec_method_online_scheduling}) to validate and execute resource scaling actions on the underlying Kubernetes cluster based on the service instances predicted in \textcolor{red}{\textbf{\large\textcircled{\small 3}}}.

\end{enumerate}

\section{The DeepScaler}
\label{sec_method}

\subsection{Performance Monitoring}
\label{sec_method_performance_monitoring}
Performance indicators provide critical information about a system, allowing developers to (i) identify and address issues that may impact the user experience, (ii) estimate resources to support the system's current and future needs, and (iii) enable effective capacity planning.
However, many cloud platforms do not offer adequate tools for monitoring applications at a highly detailed level, making it challenging to obtain accurate service-level performance indicators in real-world scenarios. 
While it is possible to achieve this through application-level instrumentation, developers must possess the necessary knowledge to expose these performance indicators. 

In this study, microservices are deployed in a Kubernetes~\cite{k8s} cluster enabled with a service mesh (i.e., Istio). With the advantages of service mesh infrastructure, \emph{DeepScaler} can easily manage microservices and collect performance indicators. 
Specifically, Istio mainly comprises a data plane and a control plane. 
The data plane is a composite entity that integrates a collection of intelligent proxies (i.e., Envoy~\cite{envoy}), which are strategically deployed as sidecars alongside microservices. These proxies adeptly mediate and control network communication between microservices. They also function as data collectors, collecting and reporting telemetry data pertinent to all mesh traffic.
The control plane manages and configures the proxies to route traffic.
The collected telemetry data is then stored in the Prometheus time-series database for querying. The data collected in our experiments are listed in Table~\ref{tab_metric}.

\begin{table}[htbp]
    \caption{Collected telemetry data and sources}
    \label{tab_metric}
    \centering
    \begin{tabular}{c}
    \hline\toprule
    \cellcolor[HTML]{ECECEC}\textbf{Istio}~\cite{istio} \textbf{\& Prometheus}~\cite{prometheus} \\
    istio\_requests\_total, istio\_request\_duration\_milliseconds\_sum, \\
    istio\_request\_duration\_milliseconds\_count\\
    \cellcolor[HTML]{ECECEC}\textbf{cAdvisor}~\cite{cadvisor} \textbf{\& Prometheus}~\cite{prometheus}\\
    container\_spec\_cpu\_period, container\_spec\_cpu\_quota,\\
    container\_cpu\_usage\_seconds\_total, container\_memory\_usage\_bytes, \\ 
    container\_spec\_memory\_limit\_bytes \\
    \bottomrule\hline
    \end{tabular}
\end{table}

\subsection{Adaptive Learning}
\begin{figure}[htbp]
    \centerline{\includegraphics[width=1\linewidth]{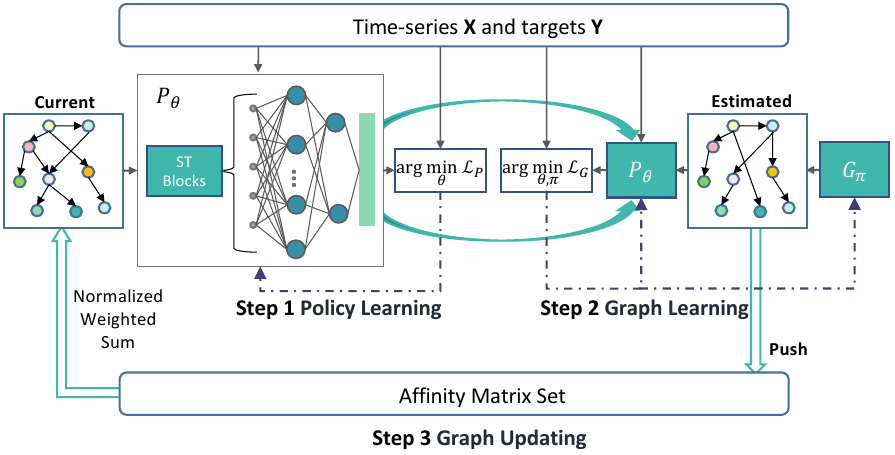}}
    \caption{Adaptive Learning Process}
    \label{fig_adaptive_learning}
\end{figure}

As shown in Fig.~\ref{fig_adaptive_learning}, the EM-based adaptive learning is an iterative process, and each iteration contains three main steps: \emph{Policy Learning}, \emph{Graph Learning}, and \emph{Graph Updating}.
\label{sec_method_adaptive_learning}

\begin{figure*}[htbp]
    \centerline{\includegraphics[width = 0.88\linewidth]{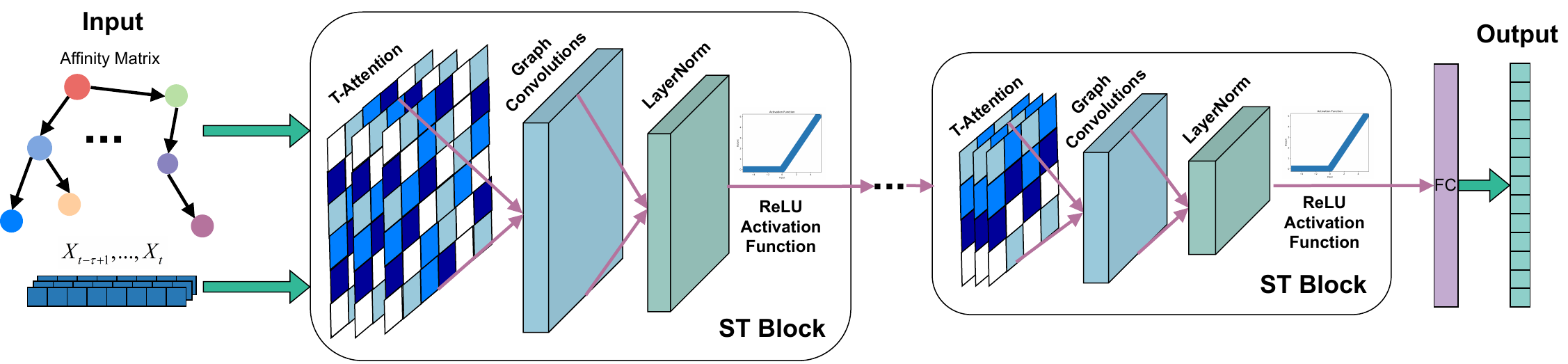}}
    \caption{Architecture of The Spatio-temporal GNN Model for Resource Estimation}
    \label{fig_model}
\end{figure*}

\textbf{Policy Learning.} \emph{Policy learning} aims to optimise the parameters of the microservice resource estimator $P$ (described in §\ref{sec_method_resource_estimation}) under the current affinity matrix of service dependency. 
The current affinity matrix $A$ is initialized by a union operation on the affinity matrix set $\mathbb{A}=\left \{ A_k|k=1,2,..., N_r \right \}$. Each $A_k\in \mathbb{R}^{N\times N}$ denotes an affinity matrix of service dependencies, and $N_r=|\mathbb{A}|$ is the total number of graphs. The $\cup$ is defined as follows: 
\begin{equation}
    \label{eq_cup}
    A^{(i,j)}=\frac{\sum_{k=1}^{N_r}A_k^{(i,j)}}{\sum_{k=1}^{N_r}\mathcal{I}[A_k^{(i,j)}]}; 
    A_k\Leftarrow \widetilde{D}_k^{-\frac{1}{2}}\widetilde{A}_k \widetilde{D}_k^{-\frac{1}{2}}
\end{equation}
where $\widetilde{D}_k^{ii}=\sum_j\widetilde{A}_k^{(ij)}$, $\widetilde{A}_k=A_k+I_N$, and $\mathcal{I}(x)$ represents the indicator function, which takes the value 1 when $x\neq0$, and 0 otherwise.

The initial $\mathbb{A}$ contains an invocation relationship between microservices automatically constructed by a distributed tracing tool (i.e., Jaeger~\cite{jaeger}). The loss function adopted in this step, represented as $\mathcal{L}_P$, is the L1 loss between the predicted sequence of instance replicas $\hat{\mathbf{Y}}$ and the ground truth $\mathbf{Y}$.

\begin{equation}
    \label{eq_p_loss}
        \mathcal{L}_P(\mathbf{Y},\hat{\mathbf{Y}})=|\mathbf{Y}-\hat{\mathbf{Y}}|_1
\end{equation}  

\textbf{Graph Learning.} \emph{Graph Learning} aims to optimize the parameters of the parameterized graph generator defined as follows:
\begin{equation} 
    \label{eq_g}
    \begin{aligned}
        &G(A)=Norm(S \odot A_{1}+(1-S) \odot A),
        \operatorname{where} \\
        A_{1}=&\operatorname{ReLU}(M_1 M_2^T-M_2 M_1^T+\operatorname{Diag}(\Lambda)), S=g(h([A_{1}, A]))  
    \end{aligned}
\end{equation}

\begin{equation}
    \label{eq_norm}
        Norm(A)=D_{2}^{-\frac{1}{2}} A_{2} D_{2}^{-\frac{1}{2}}; {A_{2}=\operatorname{ReLU}(D_{}^{-\frac{1}{2}} A_{} D_{}^{-\frac{1}{2}}-\epsilon)}
\end{equation}
where $g$ denotes a non-linear activation function (i.e., Sigmoid), $h$ is one or more $1\times 1$ convolutional layers, $\odot$ is the element-wise multiplication,
$D_2$ is diagonal matrices and $D_{2}^{(i, i)}=\sum_{j=1}^{N} A_{2}^{(i, j)}$, and $\epsilon \in (0,1)$ is a threshold used to filter out weak relationships.

The generated adjacent matrix is pushed into the affinity matrix set $\mathbb{A}$. Equations \eqref{eq_g} and \eqref{eq_norm} ensure the overall sparsity of the generated matrix, while the number of related nodes for each node is not restricted. It should be noted that some nodes may be strongly correlated with other nodes while others are relatively isolated. Therefore, the \emph{Graph Learning} can generate more effective relationships between the studied nodes.

The loss function $\mathcal{L}_G$ adopted in this step is as following:
\begin{equation}
    \label{eq_g_loss}
        \mathcal{L}_{G}(\mathbf{Y}, \hat{\mathbf{Y}})=\mathcal{L}_{P}(\mathbf{Y}, \hat{\mathbf{Y}})+\operatorname{ReLU}(\frac{C}{N^{2}}-\delta) / \delta 
\end{equation}
\begin{equation}
    \Delta A=\operatorname{ReLU}\left[\mathcal{I}\left(A_{\text {new }}\right)-\mathcal{I}\left(A_{\text {old }}\right)\right];
    C = \sum_{i=1}^{N} \sum_{j=1}^{N} \Delta A^{(i, j)}
\end{equation}
The right half of $\mathcal{L}_G$ controls the proportion of the new edge to the maximum edges of the graph by introducing a hyperparameter $\delta \in(0,1)$. 
The left half of $\mathcal{L}_G$ is $\mathcal{L}_P$, meaning that stronger correlations with the microservice resource estimation are highlighted after a few iterations, while weaker correlations are gradually erased. 

\textbf{Graph Updating.} Due to the insufficient precision and efficacy of the subgraphs within the affinity matrix set $\mathbb{A}$. We leverage prediction errors to weigh the significance of distinct subgraphs as illustrated by \eqref{eq_update_affinity_matrix}, thereby attaining an optimal estimate for matrix $A$:
\begin{equation}
    \label{eq_update_affinity_matrix}
    A=Norm(\sum_{i=1}^{N_r}\operatorname{softmax}(\max_{1\leq i \leq N_r}{L_i}-\vec{l})A_i)
\end{equation}
\begin{equation}
    \label{eq_l}
    \vec{l}=(L_{1}, \ldots, L_{N_{r}})^T;L_k = \mathcal{L}_P[P_{\theta}(\mathcal{X}|A_k),\mathbf{Y}]
\end{equation}
where $A_k \in \mathbb{A}$, and $\theta$ is the model parameter of $P$ trained in the \emph{Policy Learning} step. 

\begin{algorithm}
	\renewcommand{\algorithmicrequire}{\textbf{Input:}}
	\renewcommand{\algorithmicensure}{\textbf{Output:}}
	\caption{Pseudocode of Adaptive Learning}
	\label{alg_learning}
	\begin{algorithmic}[1]
        \REQUIRE  The input data $\mathcal{X}$, target data $\mathbf{Y}$,
        initial affinity matrix set $\mathbb{A}$ with capacity $N_{max}$,
        the GNN-based estimator $P$ with parameters $\theta$,
        the graph generator $G$ with trainable parameters $\pi$, 
       
        \ENSURE Trained $P_{\theta}$, maximum likelihood estimation of $A$
        \STATE Set $A_{\text {old }}=\bigcup_{A_i \in \mathbb{A}} A_{i}$
        \WHILE{not done}
        \STATE Train $P_{\theta}$ by minimize the loss via Adam:\\
        \ \ \ \ \ \ \ \ \ \ $\theta \leftarrow \arg\mathop{\min}\limits_{\theta}\mathcal{L}_{P}(\mathbf{Y},P(\mathcal{X},A_{old};\theta))$
        \STATE Generate new affinity matrix:\\ \ \ \ \ \  $A_{new}=G_{\pi}(A_{old})$, and push $A_{new}$ into $\mathbb{A}$
        \STATE Fix $\theta$, train $G_{\pi}$ by minimize the loss via Adam:\\
        \ \ \ \ \ \ $\pi \leftarrow \arg\mathop{\min}\limits_{\pi}\mathcal{L}_{G}(\mathbf{Y},P_{\theta}(\mathcal{X},A_{new};\pi))$
        \STATE Compute prediction loss of all subgraphs in $\mathbb{A}$ by \eqref{eq_p_loss}
        \IF{$|\mathbb{A}|>N_{\max }$}
            \STATE Remove subgraph with the max prediction loss
        \ENDIF
        \STATE Compute $A$ according to \eqref{eq_update_affinity_matrix}, and set $A_{\text {old }}=A$
        \ENDWHILE
        \STATE \Return{ $P_{\theta}$, $A$}
	\end{algorithmic}  
\end{algorithm}

\subsection{Resource Estimation Using Spatio-temporal GNN}
\label{sec_method_resource_estimation}

Accurately estimating resources and integrating with an autoscaler is crucial for the efficient implementation of autoscaling microservices, ensuring optimal operation without incurring unnecessary costs or risking performance issues.
By leveraging GNN, we accurately estimate resource needs in complex and dynamic environments, enabling efficient resource management and optimal performance with minimal overhead in autoscaling architecture.
After the rigorous collection and meticulous pre-processing of the data, the analyzed data is fed into the model, leveraging a multifaceted blend of spatial-temporal blocks that are skillfully and iteratively assembled to capture a remarkably broad and diverse spectrum of dynamic spatial-temporal relationships, seamlessly incorporating various key aspects as depicted in Figure \ref{fig_model}, which mainly includes T-Attention, Graph convolutions, LayerNorm.

\textbf{T-Attention.} T-Attention layer utilized an attention mechanism to handle the temporal dependencies among various time intervals in the workload conditions.
\begin{equation}
    \begin{gathered}
    {E}={V}_{e} \cdot \sigma((({{\mathbf{X}}}^{(r-1)})^{T} {U}_{1}) {U}_{2}({U}_{3} {{\mathbf{X}}}^{(r-1)})+{b}_{e}) \\
    {E}_{i, j}^{\prime}={\exp ({E}_{\mathrm{i}, \mathrm{j}})}/{\sum_{j=1}^{T_{r-1}} \exp ({E}_{\mathrm{i}, \mathrm{j}})}
    \end{gathered}
\end{equation}
We defined a temporal correlation matrix $E$, which semantically represents the strength of dependencies between each time $i$ and $j$. 
The calculation of $E$ is based on the varying inputs, and the learnable parameters ${V}_{e}, {b}_{e} \in {\mathbb{R}}^{T_{r-1} \times T_{r-1}}, {U}_{1} \in {\mathbb{R}}^{N}, {U}_{2} \in {\mathbb{R}}^{C_{r-1} \times N}, {U}_{3} \in {\mathbb{R}}^{C_{r-1}}$. 
$C^{r-1}$ is the number of channels of the input data and $T_{r-1}$ is the length of the temporal dimension in the $r_{th}$ layer.
The normalized temporal attention matrix $E$ is then applied directly to the input data $\mathbf{X}$ to dynamically adjust the input by merging relevant information. Equation \eqref{temporal} represents the adjusted input data after considering the temporal dependencies.
\begin{equation} \label{temporal}
\hat{{\mathbf{X}}}^{(r-1)}=({\mathbf{X}}_{1}, {\mathbf{X}}_{2}, \ldots, {\mathbf{X}}_{T_{r-1}}) {E}^{\prime} \in {\mathbb{R}}^{N \times C_{r-1} \times T_{r-1}}
\end{equation}

\textbf{Graph Convolutions.}
Graph convolutions layer captures essential relationships between microservices for network representation. We predict pod replicas based on historical data and adjacency matrices by using the ChebConv technique. This approach efficiently approximates functions and captures local graph structure information using Chebyshev polynomials. We handle irregular graph structures and reduce computational complexity by leveraging the graph spectrum. The similarity between the generated relationships ensures stability and consistency in the learning process, enabling faster convergence and effective use of prior knowledge.
\begin{equation}\label{cheb}
H^{(l)}=\operatorname{ChebConv}(\hat{A}, H^{(l-1)} ; \Theta^{(l)})
\end{equation}
As equation \eqref{cheb} shown, in the context of ChebConv, the activation function utilized is $ReLU(\sigma)$. The normalized adjacent matrix is denoted as $\hat{A}$. $H^{l}$ refers to the hidden features of the $l_{th}$ layer, while $\Theta^{(l)}$ represents the learnable parameters corresponding to the same layer.

\textbf{LayerNorm.}
LayerNorm improves convergence and reduces vulnerability to input magnitude in neural networks during training, with a mathematical formula as follows.
\begin{equation}
    LayNorm(X)=\frac{X-\mu }{\sqrt{\sigma^2+\epsilon} } \odot\gamma+\beta
\end{equation}
$X$ is the input data vector after convolution, $\mu$ and $\sigma$ are the mean and standard deviation of $X$ along each feature dimension, respectively. $\epsilon$ is a very small constant, $\odot$ denotes element-wise multiplication, and $\gamma$ and $\beta$ are learnable parameter vectors used to scale and shift the normalized data.
Subsequently, we employ $ReLU$ activation function to introduce non-linearity to the output of the previous layer and help prevent the vanishing gradient problem.

Ultimately, after feature extraction by stacked spatio-temporal blocks, the resulting tensor is fed into a fully connected (\emph{FC}) layer. The input tensor is flattened into a one-dimensional vector and undergoes a linear transformation with a matrix of learnable weights and biases. This ensures the output is given due consideration for accurate and consistent forecasting.

\subsection{Online Scheduling}
\label{sec_method_online_scheduling}

We create Docker~\cite{docker} images for each microservice in the application and save them in a private Docker hub; our images include all necessary dependencies, allowing microservices to be quickly deployed on any platform with a container engine. 
During deployment, the online scheduler uses the API of Kubernetes python client~\cite{k8sapi} to specify the number of resources the service requires at the container level based on the service instances predicted by the resource estimator.

Specifically, the Kubernetes master node receives the request for the creation instruction of a pod with specific compute resources and forwards it to the API server. The scheduler then determines the most suitable worker nodes (i.e., host servers) to run the pod and notifies the kubelet agent on the selected worker node to create the pod. Upon completion of pod creation, kubelet informs the API server that the pod is running. During runtime, the scheduler can modify the resource configuration online by submitting an update request to the master node. 
In addition, the scheduler sets a trust threshold for the prediction model. Whenever the number of prediction errors or SLA violations exceeds the threshold, the scheduler reduces trust in the model and becomes more conservative in recycling resources. 

\section{Experimental Evaluation}
\label{sec_eval}
In this section, we evaluate \emph{DeepScaler} to answer two questions:
\begin{itemize}
    \item How effective is \emph{DeepScaler}? 
    \item How adaptive is \emph{DeepScaler}?
\end{itemize}

\subsection{Experimental Settings}
\label{sec_eval_experimental_settings}

\textbf{System Setup.} The experimental evaluation was conducted on a distributed cluster deployed in a public \emph{Elastic Compute Service} (ECS) platform. The cluster contains eight virtual machines (VMs) running Ubuntu 18.04 LTS operating system with kernel version 4.15.0. Half of the VMs each have a 12-core 2.2 GHz CPU, 24 GB memory, and 100GB disk. Each of the other VMs has a 24-core 2.2 GHz CPU, 32 GB memory, and 500 GB disk. All the VMs are in the same local area network to reduce the network jitters. We used the Kubernetes container orchestration system to manage the deployment of microservices on the cluster and Istio service mesh to take over network traffic and provide load balancing.

\begin{figure}[htbp]
    \centerline{\includegraphics[width = \linewidth]{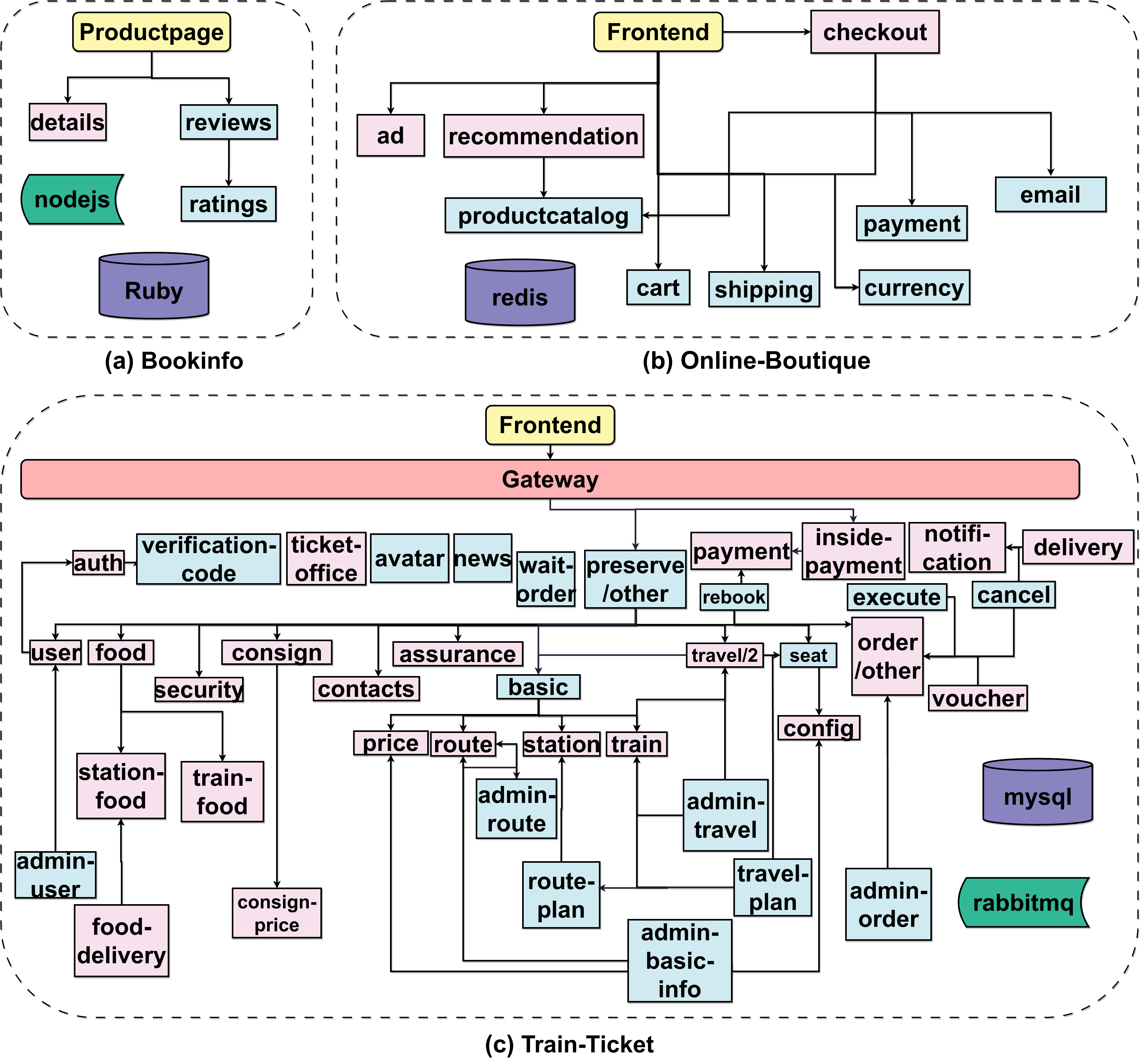}}
    \caption{Architectures of Used Benchmark Applications}
    \label{fig_benchmark}
\end{figure}

    \textbf{Benchmark Microservices.} The evaluation of \emph{DeepScaler} utilizes three end-to-end interactive and responsive real-world microservice benchmarks.
    Fig.~\ref{fig_benchmark} shows the architectures of these applications, and a short description for each is as follows:
    (i) BookInfo~\cite{bookinfo}, provided by Istio, is an online bookstore application that includes detailed information about a book, including its description, ISBN, number of pages, and reviews. The application comprises four microservices (Productpage, Reviews, Details, and Ratings), which have been incorporated to demonstrate various features of Istio.
    (ii) Online-Boutique~\cite{boutique} is a cloud-native microservices application used by Google to showcase the functionality of Kubernetes/GKE, Istio, and gRPC. This web-based e-commerce application allows users to browse a range of merchandise, add items to their cart, and complete their purchases.
    (iii) Train-Ticket~\cite{trainticket} is a ticket booking application comprising 41 microservices, each responsible for a specific function, such as user authentication, ticket booking, payment processing, and notification, for a comprehensive evaluation in a multi-functional scenario. Various programming languages are used in benchmarks, such as C\#, Java, Python, Go, Ruby, and Node.js.

\begin{figure}[htbp]
\centerline{\includegraphics[width = \linewidth]{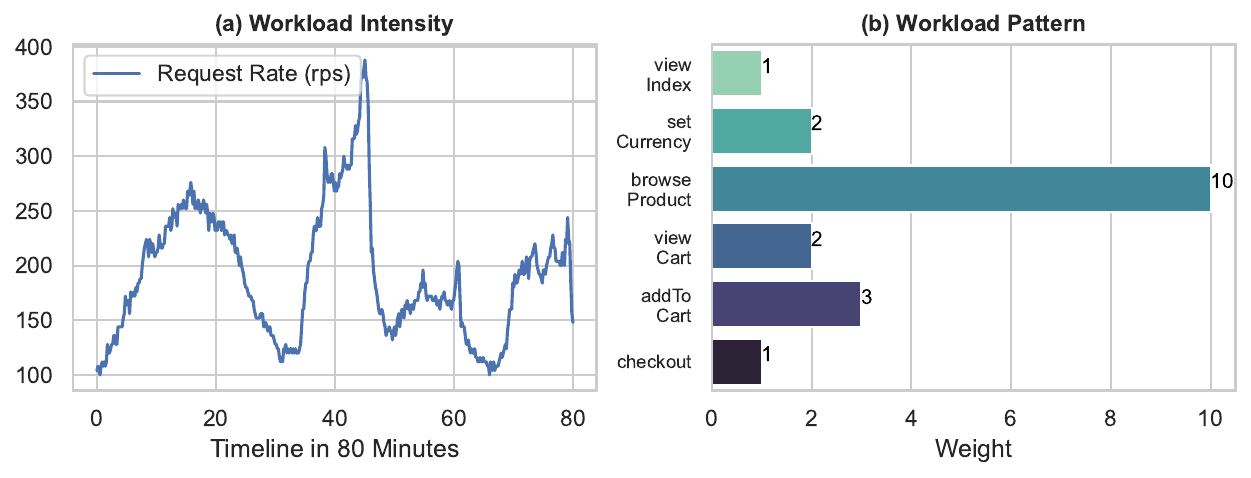}}
\caption{Benchmark Workloads}
\label{fig_eval_workloads}
\end{figure}

\textbf{Workload Generation.}
In order to replicate a live production environment, we employed Locust~\cite{locust}, an open-loop asynchronous workload generator, to drive the services. The generated workload intensity varied over time, emulating typical characteristics of microservice workloads, including slight increases, slight decreases, sharp increases, sharp decreases, and continuous fluctuations, as depicted in Fig. \ref{fig_eval_workloads}(a). All microservice benchmarks utilized this varied intensity of workloads, although each benchmark had distinct user request patterns. For instance, Online-Boutique's request pattern is presented in Fig. \ref{fig_eval_workloads}(b). To guarantee the stability of our experiments, we have deactivated all other user workloads on the cluster.

\textbf{Evaluation Metrics.}
To assess the precision of the model predictions, we employ mean absolute error (MAE), root mean square error (RMSE), and mean absolute percentage error (MAPE).
We use the SLA violation rate, resource cost, and cumulative absolute errors (CAE) to appraise the scaling policy. The violation rate and resource cost are defined by (\ref{eq_slavio}) and (\ref{eq_cost}). 
To evaluate whether \emph{DeepScaler} extracted latent service dependencies, we employ the Jaccard Similarity to compare the relationships obtained from \emph{DeepScaler} with those from origin-destination (OD) and correlation coefficient (CC) relationships. OD and CC relationships are defined as following:

\begin{equation}
    \label{od}
    \begin{aligned}
        A_{O D}^{(i, j)} & = \begin{cases}N_{p a s}(i, j) & \text { if } N_{p a s}(i, j)>{\left.\frac{\max_{j} A_{OD}^{(i, j)}}{2}\right.} \\
        0 & \text { else }\end{cases}
        \end{aligned}
\end{equation}
\begin{equation}
    \label{cc}
    A_{CC}^{(i, j)}= \begin{cases}\left|\frac{\operatorname{Cov}\left(\vec{s}_{i}, \vec{s}_{j}\right)}{\sqrt{\operatorname{var}\left(\vec{s}_{i}\right) \operatorname{var}\left(\vec{s}_{j}\right)}}\right| & \text { if }\left|\frac{\operatorname{Cov}\left(\vec{s}_{i}, \vec{s}_{j}\right)}{\sqrt{\operatorname{var}\left(\vec{s}_{i}\right) \operatorname{var}\left(\vec{s}_{j}\right)}}\right|>T C_{i} \\ 0 & \text { else }\end{cases} 
\end{equation}
where $N_{pas}(i,j)$ is the average number of requests from $S_i$ to $S_j$, $\vec{s}_{i}$ stores historical data of node $S_{i}$ and is replaced by its ranks.

\begin{figure}[htbp]
    \centerline{\includegraphics[width =\linewidth]{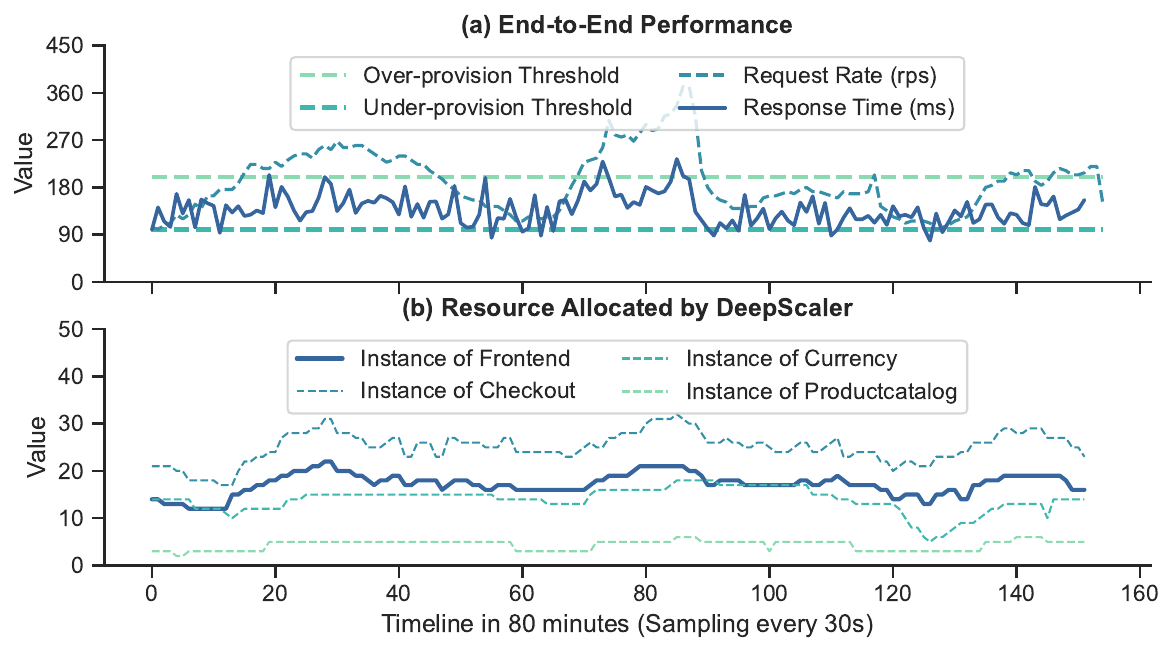}}
    \caption{A Running Example of \emph{DeepScaler}}
    \label{fig_eval_autoscaling}
\end{figure}

\begin{figure*}[htbp]
    \centerline{\includegraphics[width = \linewidth]{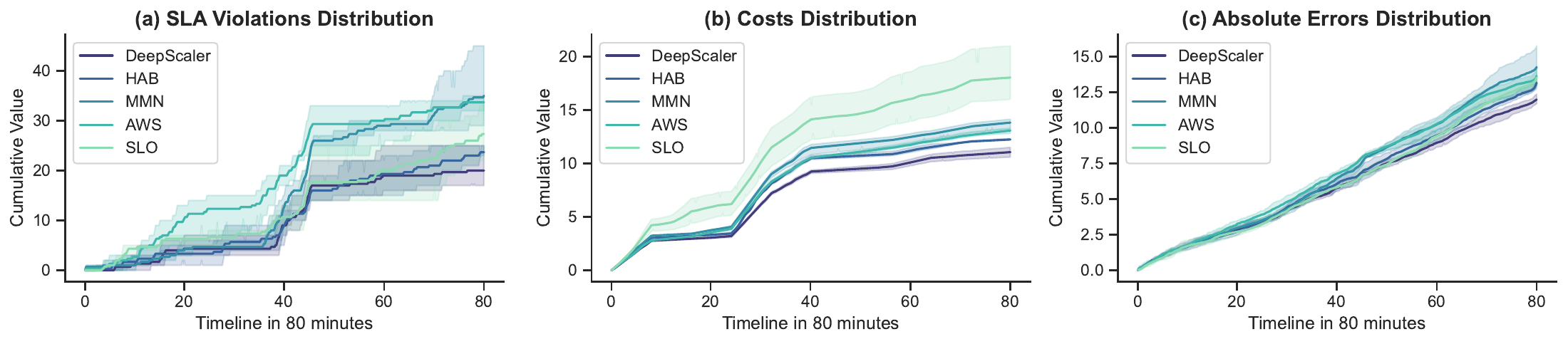}}
    \caption{Performance Comparison of Autoscaling Processes on SLA Violations, Costs, and Absolute Errors}
    \label{fig_eval_comparison}
\end{figure*}

\subsection{Overall Evaluation Of the DeepScaler (RQ1)}\label{sec_eval_overall}

\textbf{Effectiveness.} \emph{DeepScaler} performs service instance prediction every 30 seconds based on real-time telemetry indicators and scales resources accordingly.
Fig.~\ref{fig_eval_autoscaling} shows a running example of \emph{DeepScaler} autoscaling the \emph{Online Boutique} application under the benchmark workloads; we only plot the four most common services for clarity. 
From Fig.~\ref{fig_eval_autoscaling}(a), we observe that \emph{DeepScaler} controlling the vast majority of the end-to-end response time into a specific interval that neither violates SLA requirements nor over-provisioning. This fact implies that \emph{DeepScaler} achieves an effective autoscaling mechanism that guarantees SLAs while reducing resource costs. 
We attribute this to \emph{DeepScaler} correctly predicting resource needs and promptly scaling up and down the right services (as shown in Fig.~\ref{fig_eval_autoscaling}(b)).

\begin{table}[htbp]
    \caption{Autoscaling Assessment}
    \label{tab_comparison}
    \centering
    \begin{tabular}{c|c|c|c}
    \hline\toprule 
    \textbf{Auto-scaler} & \begin{tabular}{c}\textbf{Violation Rate}\\ \textit{(percent)}\end{tabular} & \begin{tabular}{c}\textbf{Cost}\\ \textit{(core-hours)}\end{tabular} & \begin{tabular}{c}\textbf{CAE}\\ \textit{(secends)}\end{tabular} \\ \cmidrule(lr){1-4}
    \emph{DeepScaler} & \textbf{3.804$\pm$0.109}  & \textbf{10.840$\pm$0.244} & \textbf{11.810$\pm$0.125} \\ 
    HAB~\cite{tong2021holistic} & 5.217$\pm$0.217 & 12.266$\pm$0.068 & 13.219$\pm$0.224 \\ 
    MMN~\cite{jiang2010autonomous} & 8.043$\pm$1.739 & 13.673$\pm$0.466 & 14.260$\pm$1.500 \\ 
    AWS~\cite{aws_autoscaling} & 7.283$\pm$0.326 & 13.066$\pm$0.252 & 13.564$\pm$0.588 \\ 
    SLO~\cite{gergin2014decentralized} & 5.978$\pm$1.196 & 18.510$\pm$2.497 & 13.581$\pm$0.409 \\ \bottomrule\hline
    \end{tabular}
\end{table}

\textbf{Comparisons.}
To compare the performance between \emph{DeepScaler} and other mainstream autoscalers, we implemented four baseline methods subjected to the same testbed as \emph{DeepScaler} to manage all service resources. These four autoscaling methods are: (i) {AWS}~\cite{aws_autoscaling} that emulates Amazon Auto-Scaling Services, (ii) {MMN}~\cite{jiang2010autonomous} based on an M/M/n/PS queuing model, (iii) {HAB}~\cite{tong2021holistic} based on a balanced queuing network, and (iv) {SLO}~\cite{gergin2014decentralized} based on SLO decomposition. 
AWS and SLO represent rule-based methods. AWS utilizes CPU utilization for rulemaking, while SLO relies on end-to-end response time. We compare these methods to assess the benefits of autoscaling using multiple indicators. HAB is a holistic autoscaling strategy considering service dependencies, while MMN is a non-holistic approach that independently models and scales. Both HAB and MMN are based on queueing theory. We contrast them to analyze the advantages of holistic autoscaling.

Table~\ref{tab_comparison} shows a comparison of the quantitative metrics of the experimental results. From the table, we find that \emph{DeepScaler} outperforms other methods in all metrics, with an average improvement of 41\% in reducing SLA violations, 23\% in reducing resource spending, and 13\% in reducing cumulative errors.
We argue that \emph{DeepScaler} outperforms other approaches because it uses an elaborate GNN that can capture spatio-temporal features of graph-structural data to model microservice resource prediction and adaptively captures latent service dependencies using a graph learning method (§\ref{sec_method_adaptive_learning}), making microservice resource prediction more accurate.
In addition, Fig.~\ref{fig_eval_comparison} shows the cumulative distribution of these metrics over time.
It can be seen that \emph{DeepScaler}'s ability to reduce both SLA violations and resource costs consistently outperforms other methods over time and becomes increasingly evident. It means that in production environments where continuous operation is required, equipping \emph{DeepScaler} not only provides better quality of service assurance but also delivers significant resource savings.

As observed in Table~\ref{tab_comparison} and Fig.~\ref{fig_eval_comparison}, the holistic autoscaling algorithm, i.e., \emph{DeepScaler} and HAB,
significantly outperforms other non-holistic autoscaling approaches. This fact implies that in microservice applications with complex service dependencies, the holistic autoscaling algorithms represent a promising solution for achieving QoS guarantees and optimal resource allocation policies. The reason is that the holistic autoscaling algorithm can scale both the bottleneck service and its associated services in one scaling action based on service dependencies, thus mitigating SLA violations as fast as possible. 
In contrast, the non-holistic approach takes more time to mitigate SLA violations.

It is also worth mentioning that at 30 to 50 minutes of the timeline axis in Fig.~\ref{fig_eval_comparison}, the bursty workloads exhibit higher SLA violations, although all autoscaling methods have allocated more resources. That is because burst workloads change rapidly and have unpredictable emergence times and request volumes, challenging prediction-based proactive and rule-based reactive approaches. Therefore, studying further fine-grained handling of workload bursts is vital to improving the performance of all dynamic workload-oriented resource management algorithms.

\begin{figure}[htbp]
    \centerline{\includegraphics[width = \linewidth]{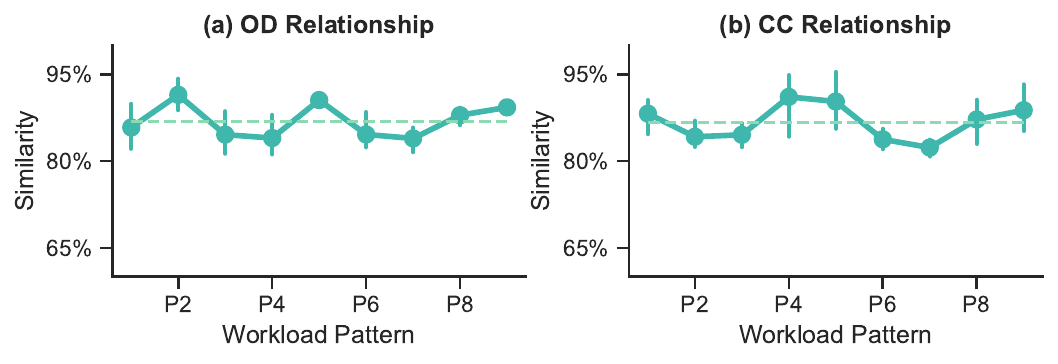}}
    \caption{Dependency Capture Capability on Different Workload Patterns}
    \label{fig_eval_similarity}
\end{figure}

\begin{figure*}[htbp]
    \centerline{\includegraphics[width = \linewidth]{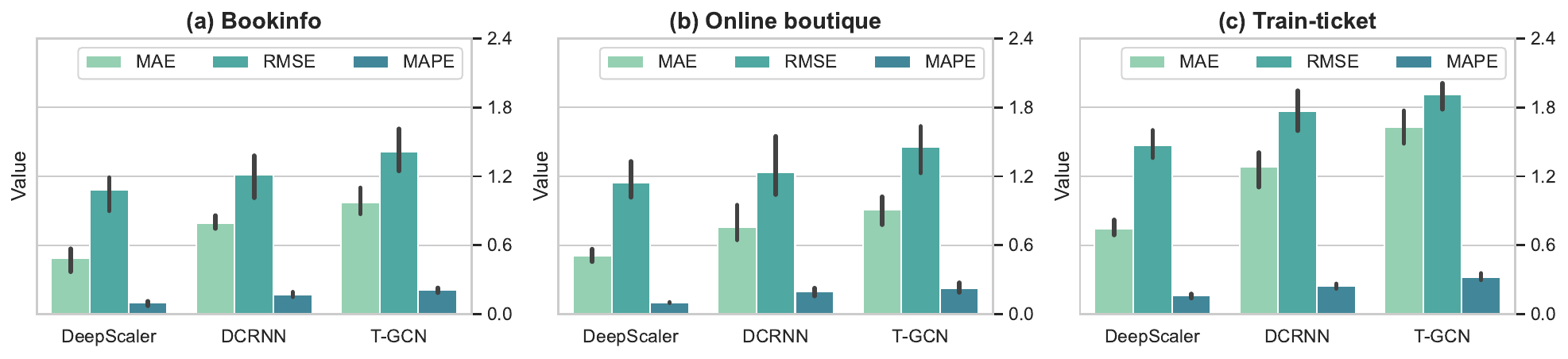}}
    \caption{Performance Comparison of Resource Estimations on Different Applications}
    \label{fig_eval_resource_estimation}
\end{figure*}

\subsection{Adaptability Evaluation (RQ2)}
\label{sec_adaptive_evaluation}

\textbf{Varying Different Workload Patterns.}
In practice, the workload intensity and patterns of microservices usually vary over time. §\ref{sec_eval_overall} has verified that \emph{DeepScaler} can efficiently allocate resources under a specific pattern and varying intensity workload (as shown in Fig.~\ref{fig_eval_workloads}). Here, we verify its adaptability to different workload patterns by tuning different weights of user request types. Fig.~\ref{fig_eval_similarity} shows the accuracy of \emph{DeepScaler} in capturing service dependencies at different weights. From the figure, we observe that the accuracy in capturing OD and CC relationships always remains in an acceptable range (i.e., 89\%$\pm$7\%). It illustrates the adaptability of \emph{DeepScaler} to workload pattern changes. Note that the initial affinity matrices of \emph{DeepScaler} contain only invocation relationships. Therefore, the accurate capture of OD and CC relationships by \emph{DeepScaler} validates its ability to adaptively capture latent service dependencies.

\textbf{Varying Different Applications.} To validate the adaptability of \emph{DeepScaler} to different microservices and to evaluate the impact of potential service dependencies, we additionally implemented two GNN-based baseline algorithms: (i) a diffusion convolution recurrent neural network (DCRNN)~\cite{li2017diffusion} and (ii) a temporal graph convolutional network (T-GCN)~\cite{zhao2019t} to perform resource prediction for the three different benchmark applications (§\ref{sec_eval_experimental_settings}).
Fig.~\ref{fig_eval_resource_estimation} displays the experimental results. 
From the figure, we find that although the prediction error of all algorithms gradually increases as the size of microservices increases, \emph{DeepScaler} has the lowest error growth rate compared with other methods.
That means \emph{DeepScaler} has better adaptability to varying microservice application sizes and architectures. That is to say, \emph{DeepScaler} is adaptive in managing the resources of large-scale microservices. The reason is that other GNNs only use predefined affinity matrices (i.e., invocation relationships), while \emph{DeepScaler} is able to capture more potential service dependencies through an adaptive graph learning approach, thus enabling better adaptation to more complex microservice scenarios.

\section{Discussion}
\textbf{Alternate Design Choices.} 
In our study, we employed a temporal attention-based graph convolutional network, a GNN, capturing spatiotemporal features of historical data to estimate the resource needs of microservices, and an EM-based graph learning module to discover service dependencies adaptively. While our proposed approach demonstrates effectiveness in achieving the desired outcomes, a rich landscape of spatiotemporal GNNs and graph learning methods could have been explored for similar purposes. 
For instance, alternative spatiotemporal GNNs, such as STFGNN~\cite{Li_Zhu_2021}, GraphSAGE~\cite{NIPS2017_5dd9db5e}, or GAT~\cite{velickovic2017graph}, could have been contemplated to apprehend diverse dimensions of the underlying spatiotemporal features within the data.
Alternative graph learning techniques, encompassing the GRU of STGNN~\cite{STGNN}, STGSA~\cite{STGSA}, and GMAN~\cite{Zheng_Fan_Wang_Qi_2020}, among others, could have been employed to unearth service dependencies in a varied manner.
By delving into this rich landscape of possibilities, we can further enhance resource estimation's versatility, accuracy, and applicability for microservices autoscaling in various dynamic and evolving environments.

\textbf{Overhead.} Table~\ref{tab_overhead} shows the overhead of \emph{DeepScaler}. \emph{Performance Monitoring} requires a resident time-series database (i.e., Prometheus) with an overhead that includes collecting and storing all performance indicators, which is about 64m core of CPU and 716 MB of memory.
\emph{Adaptive Learning} uses collected data to train the resource prediction model, which consumes 10.64 cores and takes 158.8 seconds. The retraining interval depends on whether changes in the microservice environment result in an unacceptable autoscaling performance.
\emph{DeepScaler} then performs \emph{Resource Estimation} and corresponding \emph{Online Scheduling} every 1 minute, consuming 0.107 cores of CPU and 232 MB of memory each time. In summary, \emph{DeepScaler}'s overhead primarily comes from \emph{Performance Monitoring}, which is essential for all autoscaling algorithms. Compared to the resources it can save and the benefits of QoS assurance, \emph{DeepScaler}'s overhead is negligible.

\begin{table}[htbp]
    \center
    \caption{The Overhead of DeepScaler}
    \label{tab_overhead}
    \begin{tabular}{lccc}
    \hline\toprule
    \multicolumn{1}{c}{\multirow{4}{*}{One-shot Operation}} & \multicolumn{3}{c}{Overhead} \\ \cmidrule(l){2-4} 
    \multicolumn{1}{c}{} & \multicolumn{1}{c}{\begin{tabular}{c}{CPU}\\ \textit{(core)}\end{tabular}} & \multicolumn{1}{c}{\begin{tabular}{c}{Mem}\\ \textit{(MB)}\end{tabular}} & \multicolumn{1}{c}{\begin{tabular}{c}{Time}\\ \textit{(s)}\end{tabular}} \\ \midrule
    Performance Monitoring & 64$\pm$23m  & 716$\pm$54 & $+\infty$             \\
    Adaptive Learning      & 10640$\pm$300m & 2108$\pm$212 & 158.8$\pm$2.1     \\
    Estimation \& Scheduling    & 107$\pm$8m  & 232$\pm$34 & 0.5$\pm$ 0.0  \\
    \bottomrule\hline
    \end{tabular}
\end{table}

\textbf{Limitation.}
First, \emph{DeepScaler} is a data-driven approach limited by the quality and quantity of available historical data. Second, while \emph{DeepScaler} can effectively mitigate SLA violations caused by workload fluctuations, it cannot detect and mitigate SLA violations caused by other system anomalies (e.g., network congestion, hardware failures, etc.).
Finally, the current study focused on autoscaling microservices in a container-based cloud. How well \emph{DeepScaler} would perform in other types of clouds needs to be clarified.

\section{RELATED WORK}
Autoscaling techniques have garnered substantial attention as a means to dynamically manage the allocation of computing resources in response to varying workload demands. This section comprehensively reviews autoscaling techniques, categorized into distinct classes based on their underlying methodologies.

\textbf{Rule-based} autoscaling methods employ pre-defined thresholds and heuristics to trigger scaling actions~\cite{gergin2014decentralized,VAYGHAN2021110924,SRIRAMA2020102629}. Their simplicity and ease of implementation characterize these methods. Commonly used rules include CPU utilization thresholds, response time limits, and queue length triggers. 
Most cloud platforms, such as Google Cloud and Amazon Web Service, use this technique~\cite{aws_autoscaling}.
While straightforward, rule-based approaches may struggle to accommodate intricate workload dynamics, adapt to rapidly changing conditions, and optimize resource allocation under complex scenarios. 

\textbf{Model-based} autoscaling methods employ formal mathematical models to establish relationships between workload characteristics and resource provisioning~\cite{YAN2021107216,Wang2021,fourati2022epma}. Queuing theory~\cite{jiang2010autonomous, tong2021holistic}, network models~\cite{mekki2022microservices}, and control theory~\cite{fodor2022lsso} are frequently utilized in this category. These models provide a theoretical foundation for predicting system behavior and optimizing resource allocation. However, model-based approaches rely heavily on accurate system dynamics assumptions and may struggle to capture real-world intricacies and deviations.

\textbf{Reinforcement Learning-based} autoscaling paradigms empower systems to autonomously learn optimal scaling policies by iteratively exploring interactions with their environment~\cite{xu2022coscal,Chen2023,karypiadis2022scal,Rossi2020,Khaleq2021}.
For example, AutoMan et al.~\cite{cai2023automan} propose a resource allocation method using multi-agent deep deterministic policy gradient in reinforcement learning to meet end-to-end tail latency SLOs for microservices.
While inherently adaptive, these approaches often require substantial computational resources for training and necessitate careful consideration of exploration-exploitation trade-offs.

\textbf{Machine Learning-based} autoscaling techniques encompass a spectrum of data-driven methods, including supervised and unsupervised learning, to discern patterns and correlations within historical data~\cite{song2022automatic,baarzi2021showar,li2021joint}. 
For example, Sage et al.~\cite{zhang2021sinan} used unsupervised ML models to avoid the cost of tracking tags, capture the impact of dependencies between microservices, and apply corrective measures to restore the QoS of cloud services.
Li et al.~\cite{li2023topology} optimized the network overhead of microservice applications by utilizing the topology of microservices and clusters through a heuristic graph mapping algorithm.
Compared to our work, these autoscaling methods superficially introduce service dependencies without dynamically adaptive modeling from a spatiotemporal perspective.

\section{Conclusion and Future Work}
In this paper, we present \emph{DeepScaler}, a deep learning-based holistic autoscaling approach to manage resource allocation of containerized microservices under dynamic workloads to optimize SLA assurance and cost efficiency. With an elaborate attention-based GNN model and an adaptive graph learning method, \emph{DeepScaler} can capture well microservices' spatio-temporal features and latent service dependencies to accurately estimate and proactively provision resources. Experiments validate the effectiveness and adaptability of \emph{DeepScaler}. Compared with leading autoscaling approaches, \emph{DeepScaler} obtains a more efficient resource allocation strategy and significantly reduces SLA violations at a lower cost. 
In the future, we intend to explore holistic autoscaling in extremely heterogeneous and geographically distributed computing resources to improve SLA guarantees for microservices under hybrid clouds. 
The source code of \emph{DeepScaler} is available at \url{https://github.com/SYSU-Workflow-Lab/DeepScaler}.

\section*{Acknowledgements}
We express our sincere gratitude to the reviewers for their insightful comments and constructive suggestions, which significantly contributed to the refinement of this paper. 
This research is supported by the NSFC-Guangdong Joint Fund Project (Grant No. U20A6003), the National Natural Science Foundation of China (NSFC) (Grant No. 61972427), and the Research Foundation of Science and Technology Plan Project in Guangdong Province (Grant No. 2020A0505100030).

\bibliographystyle{IEEEtran}
\bibliography{IEEEabrv,reference,mybibliography}

\begin{thebibliography}{10}
\providecommand{\url}[1]{#1}
\csname url@samestyle\endcsname
\providecommand{\newblock}{\relax}
\providecommand{\bibinfo}[2]{#2}
\providecommand{\BIBentrySTDinterwordspacing}{\spaceskip=0pt\relax}
\providecommand{\BIBentryALTinterwordstretchfactor}{4}
\providecommand{\BIBentryALTinterwordspacing}{\spaceskip=\fontdimen2\font plus
\BIBentryALTinterwordstretchfactor\fontdimen3\font minus
  \fontdimen4\font\relax}
\providecommand{\BIBforeignlanguage}[2]{{%
\expandafter\ifx\csname l@#1\endcsname\relax
\typeout{** WARNING: IEEEtran.bst: No hyphenation pattern has been}%
\typeout{** loaded for the language `#1'. Using the pattern for}%
\typeout{** the default language instead.}%
\else
\language=\csname l@#1\endcsname
\fi
#2}}
\providecommand{\BIBdecl}{\relax}
\BIBdecl

\bibitem{coutinho2015elasticity}
E.~F. Coutinho, F.~R. de~Carvalho~Sousa, P.~A.~L. Rego, D.~G. Gomes, and J.~N.
  de~Souza, ``Elasticity in cloud computing: a survey,'' \emph{annals of
  telecommunications-annales des t{\'e}l{\'e}communications}, vol.~70, pp.
  289--309, 2015.

\bibitem{chen2016self}
T.~Chen and R.~Bahsoon, ``Self-adaptive and online qos modeling for cloud-based
  software services,'' \emph{IEEE Transactions on Software Engineering},
  vol.~43, no.~5, pp. 453--475, 2017.

\bibitem{al2017elasticity}
Y.~Al-Dhuraibi, F.~Paraiso, N.~Djarallah, and P.~Merle, ``Elasticity in cloud
  computing: State of the art and research challenges,'' \emph{IEEE
  Transactions on Services Computing}, vol.~11, no.~2, pp. 430--447, 2018.

\bibitem{chen2018survey}
\BIBentryALTinterwordspacing
T.~Chen, R.~Bahsoon, and X.~Yao, ``A survey and taxonomy of self-aware and
  self-adaptive cloud autoscaling systems,'' \emph{ACM Comput. Surv.}, vol.~51,
  no.~3, jun 2018. [Online]. Available: \url{https://doi.org/10.1145/3190507}
\BIBentrySTDinterwordspacing

\bibitem{qu2018auto}
\BIBentryALTinterwordspacing
C.~Qu, R.~N. Calheiros, and R.~Buyya, ``Auto-scaling web applications in
  clouds: A taxonomy and survey,'' \emph{ACM Comput. Surv.}, vol.~51, no.~4,
  jul 2018. [Online]. Available: \url{https://doi.org/10.1145/3148149}
\BIBentrySTDinterwordspacing

\bibitem{zhong2022machine}
\BIBentryALTinterwordspacing
Z.~Zhong, M.~Xu, M.~A. Rodriguez, C.~Xu, and R.~Buyya, ``Machine learning-based
  orchestration of containers: A taxonomy and future directions,'' \emph{ACM
  Comput. Surv.}, vol.~54, no. 10s, sep 2022. [Online]. Available:
  \url{https://doi.org/10.1145/3510415}
\BIBentrySTDinterwordspacing

\bibitem{dragoni2017microservices}
N.~Dragoni, S.~Giallorenzo, A.~L. Lafuente, M.~Mazzara, F.~Montesi,
  R.~Mustafin, and L.~Safina, ``Microservices: yesterday, today, and
  tomorrow,'' \emph{Present and ulterior software engineering}, pp. 195--216,
  2017.

\bibitem{yang2019miras}
Z.~Yang, P.~Nguyen, H.~Jin, and K.~Nahrstedt, ``Miras: Model-based
  reinforcement learning for microservice resource allocation over scientific
  workflows,'' in \emph{2019 IEEE 39th International Conference on Distributed
  Computing Systems (ICDCS)}, 2019, pp. 122--132.

\bibitem{gan2019open}
\BIBentryALTinterwordspacing
Y.~Gan, Y.~Zhang, D.~Cheng, A.~Shetty, P.~Rathi, N.~Katarki, A.~Bruno, J.~Hu,
  B.~Ritchken, B.~Jackson, K.~Hu, M.~Pancholi, Y.~He, B.~Clancy, C.~Colen,
  F.~Wen, C.~Leung, S.~Wang, L.~Zaruvinsky, M.~Espinosa, R.~Lin, Z.~Liu,
  J.~Padilla, and C.~Delimitrou, ``An open-source benchmark suite for
  microservices and their hardware-software implications for cloud \& edge
  systems,'' in \emph{Proceedings of the Twenty-Fourth International Conference
  on Architectural Support for Programming Languages and Operating Systems},
  ser. ASPLOS '19.\hskip 1em plus 0.5em minus 0.4em\relax New York, NY, USA:
  Association for Computing Machinery, 2019, p. 3–18. [Online]. Available:
  \url{https://doi.org/10.1145/3297858.3304013}
\BIBentrySTDinterwordspacing

\bibitem{singh2019research}
P.~Singh, P.~Gupta, K.~Jyoti, and A.~Nayyar, ``Research on auto-scaling of web
  applications in cloud: survey, trends and future directions,'' \emph{Scalable
  Computing: Practice and Experience}, vol.~20, no.~2, pp. 399--432, 2019.

\bibitem{DARADKEH2023102713}
\BIBentryALTinterwordspacing
T.~Daradkeh and A.~Agarwal, ``Modeling and optimizing micro-service based cloud
  elastic management system,'' \emph{Simulation Modelling Practice and Theory},
  vol. 123, p. 102713, 2023. [Online]. Available:
  \url{https://www.sciencedirect.com/science/article/pii/S1569190X22001824}
\BIBentrySTDinterwordspacing

\bibitem{mirhosseini2021parslo}
\BIBentryALTinterwordspacing
A.~Mirhosseini, S.~Elnikety, and T.~F. Wenisch, ``Parslo: A gradient
  descent-based approach for near-optimal partial slo allotment in
  microservices,'' in \emph{Proceedings of the ACM Symposium on Cloud
  Computing}, ser. SoCC '21.\hskip 1em plus 0.5em minus 0.4em\relax New York,
  NY, USA: Association for Computing Machinery, 2021, p. 442–457. [Online].
  Available: \url{https://doi.org/10.1145/3472883.3486985}
\BIBentrySTDinterwordspacing

\bibitem{horovitz2018efficient}
S.~Horovitz and Y.~Arian, ``Efficient cloud auto-scaling with sla objective
  using q-learning,'' in \emph{2018 IEEE 6th International Conference on Future
  Internet of Things and Cloud (FiCloud)}, 2018, pp. 85--92.

\bibitem{urgaonkar2008agile}
\BIBentryALTinterwordspacing
B.~Urgaonkar, P.~Shenoy, A.~Chandra, P.~Goyal, and T.~Wood, ``Agile dynamic
  provisioning of multi-tier internet applications,'' \emph{ACM Trans. Auton.
  Adapt. Syst.}, vol.~3, no.~1, mar 2008. [Online]. Available:
  \url{https://doi.org/10.1145/1342171.1342172}
\BIBentrySTDinterwordspacing

\bibitem{zhang2020sarsa}
S.~Zhang, T.~Wu, M.~Pan, C.~Zhang, and Y.~Yu, ``A-sarsa: A predictive container
  auto-scaling algorithm based on reinforcement learning,'' in \emph{2020 IEEE
  International Conference on Web Services (ICWS)}, 2020, pp. 489--497.

\bibitem{zafeiropoulos2022reinforcement}
\BIBentryALTinterwordspacing
A.~Zafeiropoulos, E.~Fotopoulou, N.~Filinis, and S.~Papavassiliou,
  ``Reinforcement learning-assisted autoscaling mechanisms for serverless
  computing platforms,'' \emph{Simulation Modelling Practice and Theory}, vol.
  116, p. 102461, 2022. [Online]. Available:
  \url{https://www.sciencedirect.com/science/article/pii/S1569190X21001507}
\BIBentrySTDinterwordspacing

\bibitem{cai2023automan}
\BIBentryALTinterwordspacing
B.~Cai, B.~Wang, M.~Yang, and Q.~Guo, ``Automan: Resource-efficient
  provisioning with tail latency guarantees for microservices,'' \emph{Future
  Generation Computer Systems}, vol. 143, pp. 61--75, 2023. [Online].
  Available:
  \url{https://www.sciencedirect.com/science/article/pii/S0167739X23000213}
\BIBentrySTDinterwordspacing

\bibitem{delimitrou2016hcloud}
\BIBentryALTinterwordspacing
C.~Delimitrou and C.~Kozyrakis, ``Hcloud: Resource-efficient provisioning in
  shared cloud systems,'' \emph{SIGARCH Comput. Archit. News}, vol.~44, no.~2,
  p. 473–488, mar 2016. [Online]. Available:
  \url{https://doi.org/10.1145/2980024.2872365}
\BIBentrySTDinterwordspacing

\bibitem{delimitrou2015tarcil}
\BIBentryALTinterwordspacing
C.~Delimitrou, D.~Sanchez, and C.~Kozyrakis, ``Tarcil: Reconciling scheduling
  speed and quality in large shared clusters,'' in \emph{Proceedings of the
  Sixth ACM Symposium on Cloud Computing}, ser. SoCC '15.\hskip 1em plus 0.5em
  minus 0.4em\relax New York, NY, USA: Association for Computing Machinery,
  2015, p. 97–110. [Online]. Available:
  \url{https://doi.org/10.1145/2806777.2806779}
\BIBentrySTDinterwordspacing

\bibitem{microservicesWorkshop}
\BIBentryALTinterwordspacing
A.~Cockcroft. (2016) Microservices workshop: Why, what, and how to get there.
  Accessed: Aug 10, 2023. [Online]. Available:
  \url{https://www.slideshare.net/adriancockcroft/microservices-workshop-craft-conference}
\BIBentrySTDinterwordspacing

\bibitem{evolutionMicroservices}
\BIBentryALTinterwordspacing
------. (2016) The evolution of microservices. Accessed: Aug 10, 2023.
  [Online]. Available:
  \url{https://www.slideshare.net/adriancockcroft/evolution-of-microservices-craft-conference}
\BIBentrySTDinterwordspacing

\bibitem{grpc}
\BIBentryALTinterwordspacing
(2023) grpc. Accessed: Aug 10, 2023. [Online]. Available:
  \url{https://grpc.io/}
\BIBentrySTDinterwordspacing

\bibitem{richardson2008restful}
L.~Richardson and S.~Ruby, \emph{RESTful web services}.\hskip 1em plus 0.5em
  minus 0.4em\relax " O'Reilly Media, Inc.", 2008.

\bibitem{zhou2020graph}
\BIBentryALTinterwordspacing
J.~Zhou, G.~Cui, S.~Hu, Z.~Zhang, C.~Yang, Z.~Liu, L.~Wang, C.~Li, and M.~Sun,
  ``Graph neural networks: A review of methods and applications,'' \emph{AI
  Open}, vol.~1, pp. 57--81, 2020. [Online]. Available:
  \url{https://www.sciencedirect.com/science/article/pii/S2666651021000012}
\BIBentrySTDinterwordspacing

\bibitem{wu2020comprehensive}
Z.~Wu, S.~Pan, F.~Chen, G.~Long, C.~Zhang, and P.~S. Yu, ``A comprehensive
  survey on graph neural networks,'' \emph{IEEE Transactions on Neural Networks
  and Learning Systems}, vol.~32, no.~1, pp. 4--24, 2021.

\bibitem{zhang2022deep}
Z.~Zhang, P.~Cui, and W.~Zhu, ``Deep learning on graphs: A survey,'' \emph{IEEE
  Transactions on Knowledge and Data Engineering}, vol.~34, no.~1, pp.
  249--270, 2022.

\bibitem{sankar2021graph}
\BIBentryALTinterwordspacing
A.~Sankar, Y.~Liu, J.~Yu, and N.~Shah, ``Graph neural networks for friend
  ranking in large-scale social platforms,'' in \emph{Proceedings of the Web
  Conference 2021}, ser. WWW '21.\hskip 1em plus 0.5em minus 0.4em\relax New
  York, NY, USA: Association for Computing Machinery, 2021, p. 2535–2546.
  [Online]. Available: \url{https://doi.org/10.1145/3442381.3450120}
\BIBentrySTDinterwordspacing

\bibitem{chen2017supervised}
Z.~Chen, X.~Li, and J.~Bruna, ``Supervised community detection with line graph
  neural networks,'' \emph{arXiv preprint arXiv:1705.08415}, 2017.

\bibitem{trainticket}
\BIBentryALTinterwordspacing
(2022) {Train Ticket: A Benchmark Microservice System}. Accessed: Aug 10, 2023.
  [Online]. Available: \url{https://github.com/FudanSELab/train-ticket}
\BIBentrySTDinterwordspacing

\bibitem{boutique}
\BIBentryALTinterwordspacing
(2023) {Online Boutique: A microservice benchmark used by Google Cloud
  Platform}. Accessed: Aug 10, 2023. [Online]. Available:
  \url{https://github.com/GoogleCloudPlatform/microservices-demo}
\BIBentrySTDinterwordspacing

\bibitem{bookinfo}
\BIBentryALTinterwordspacing
(2023) {Bookinfo: A microservice benchmark provided by Istio}. Accessed: Aug
  10, 2023. [Online]. Available: \url{https://istio.io/docs/examples/bookinfo/}
\BIBentrySTDinterwordspacing

\bibitem{yu2019microscaler}
G.~Yu, P.~Chen, and Z.~Zheng, ``Microscaler: Automatic scaling for
  microservices with an online learning approach,'' in \emph{2019 IEEE
  International Conference on Web Services (ICWS)}.\hskip 1em plus 0.5em minus
  0.4em\relax IEEE, 2019, pp. 68--75.

\bibitem{jiang2010autonomous}
\BIBentryALTinterwordspacing
D.~Jiang, G.~Pierre, and C.-H. Chi, ``Autonomous resource provisioning for
  multi-service web applications,'' in \emph{Proceedings of the 19th
  International Conference on World Wide Web}, ser. WWW '10.\hskip 1em plus
  0.5em minus 0.4em\relax New York, NY, USA: Association for Computing
  Machinery, 2010, p. 471–480. [Online]. Available:
  \url{https://doi.org/10.1145/1772690.1772739}
\BIBentrySTDinterwordspacing

\bibitem{meng2022hra}
C.~Meng, J.~Tong, M.~Pan, and Y.~Yu, ``Hra: An intelligent holistic resource
  autoscaling framework for multi-service applications,'' in \emph{2022 IEEE
  International Conference on Web Services (ICWS)}, 2022, pp. 129--139.

\bibitem{tong2021holistic}
J.~Tong, M.~Wei, M.~Pan, and Y.~Yu, ``A holistic auto-scaling algorithm for
  multi-service applications based on balanced queuing network,'' in \emph{2021
  IEEE International Conference on Web Services (ICWS)}, 2021, pp. 531--540.

\bibitem{yu2020microscaler}
G.~Yu, P.~Chen, and Z.~Zheng, ``Microscaler: Cost-effective scaling for
  microservice applications in the cloud with an online learning approach,''
  \emph{IEEE Transactions on Cloud Computing}, vol.~10, no.~2, pp. 1100--1116,
  2022.

\bibitem{Quality}
C.~Delimitrou and C.~Kozyrakis, ``Quality-of-service-aware scheduling in
  heterogeneous data centers with paragon,'' \emph{IEEE Micro}, vol.~34, no.~3,
  pp. 17--30, 2014.

\bibitem{k8s}
\BIBentryALTinterwordspacing
(2023) Kubernetes: Production-grade container orchestration. Accessed: Aug 10,
  2023. [Online]. Available: \url{https://kubernetes.io/}
\BIBentrySTDinterwordspacing

\bibitem{envoy}
\BIBentryALTinterwordspacing
(2023) Envoy proxy. Accessed: Aug 10, 2023. [Online]. Available:
  \url{https://www.envoyproxy.io/}
\BIBentrySTDinterwordspacing

\bibitem{istio}
\BIBentryALTinterwordspacing
(2023) {Istio: An open source service mesh that layers transparently onto
  existing distributed applications}. Accessed: Aug 10, 2023. [Online].
  Available: \url{https://istio.io/}
\BIBentrySTDinterwordspacing

\bibitem{prometheus}
\BIBentryALTinterwordspacing
(2023) The prometheus monitoring system and time series database. Accessed: Aug
  10, 2023. [Online]. Available: \url{https://github.com/prometheus/prometheus}
\BIBentrySTDinterwordspacing

\bibitem{cadvisor}
\BIBentryALTinterwordspacing
(2023) cadvisor: Container advisor. Accessed: Aug 10, 2023. [Online].
  Available: \url{https://github.com/google/cadvisor}
\BIBentrySTDinterwordspacing

\bibitem{jaeger}
\BIBentryALTinterwordspacing
{Cloud Native Computing Foundation}. (2023) Jaeger: Distributed tracing system.
  Accessed: Aug 10, 2023. [Online]. Available:
  \url{https://www.jaegertracing.io/}
\BIBentrySTDinterwordspacing

\bibitem{docker}
\BIBentryALTinterwordspacing
{Docker, Inc.} (2023) Docker: Open platform for developing, shipping, and
  running applications. Accessed: Aug 10, 2023. [Online]. Available:
  \url{https://www.docker.com/}
\BIBentrySTDinterwordspacing

\bibitem{k8sapi}
\BIBentryALTinterwordspacing
(2023) Python client for the kubernetes api. Accessed: Aug 10, 2023. [Online].
  Available: \url{https://github.com/kubernetes-client/python}
\BIBentrySTDinterwordspacing

\bibitem{locust}
\BIBentryALTinterwordspacing
(2023) Locust performance testing tool. Accessed: Aug 10, 2023. [Online].
  Available: \url{https://locust.io/}
\BIBentrySTDinterwordspacing

\bibitem{aws_autoscaling}
\BIBentryALTinterwordspacing
(2023) {Amazon Auto Scaling Service}. Accessed: Aug 10, 2023. [Online].
  Available: \url{http://aws.amazon.com/autoscaling/}
\BIBentrySTDinterwordspacing

\bibitem{gergin2014decentralized}
I.~Gergin, B.~Simmons, and M.~Litoiu, ``A decentralized autonomic architecture
  for performance control in the cloud,'' in \emph{2014 IEEE International
  Conference on Cloud Engineering}, 2014, pp. 574--579.

\bibitem{li2017diffusion}
Y.~Li, R.~Yu, C.~Shahabi, and Y.~Liu, ``Diffusion convolutional recurrent
  neural network: Data-driven traffic forecasting,'' \emph{arXiv preprint
  arXiv:1707.01926v3}, 2018.

\bibitem{zhao2019t}
L.~Zhao, Y.~Song, C.~Zhang, Y.~Liu, P.~Wang, T.~Lin, M.~Deng, and H.~Li,
  ``T-gcn: A temporal graph convolutional network for traffic prediction,''
  \emph{IEEE Transactions on Intelligent Transportation Systems}, vol.~21,
  no.~9, pp. 3848--3858, 2020.

\bibitem{Li_Zhu_2021}
\BIBentryALTinterwordspacing
M.~Li and Z.~Zhu, ``Spatial-temporal fusion graph neural networks for traffic
  flow forecasting,'' \emph{Proceedings of the AAAI Conference on Artificial
  Intelligence}, vol.~35, no.~5, pp. 4189--4196, May 2021. [Online]. Available:
  \url{https://ojs.aaai.org/index.php/AAAI/article/view/16542}
\BIBentrySTDinterwordspacing

\bibitem{NIPS2017_5dd9db5e}
\BIBentryALTinterwordspacing
W.~Hamilton, Z.~Ying, and J.~Leskovec, ``Inductive representation learning on
  large graphs,'' in \emph{Advances in Neural Information Processing Systems},
  I.~Guyon, U.~V. Luxburg, S.~Bengio, H.~Wallach, R.~Fergus, S.~Vishwanathan,
  and R.~Garnett, Eds., vol.~30.\hskip 1em plus 0.5em minus 0.4em\relax Curran
  Associates, Inc., 2017. [Online]. Available:
  \url{https://proceedings.neurips.cc/paper_files/paper/2017/file/5dd9db5e033da9c6fb5ba83c7a7ebea9-Paper.pdf}
\BIBentrySTDinterwordspacing

\bibitem{velickovic2017graph}
P.~Velickovic, G.~Cucurull, A.~Casanova, A.~Romero, P.~Lio, Y.~Bengio
  \emph{et~al.}, ``Graph attention networks,'' \emph{stat}, vol. 1050, no.~20,
  pp. 10--48\,550, 2017.

\bibitem{STGNN}
\BIBentryALTinterwordspacing
X.~Wang, Y.~Ma, Y.~Wang, W.~Jin, X.~Wang, J.~Tang, C.~Jia, and J.~Yu, ``Traffic
  flow prediction via spatial temporal graph neural network,'' in
  \emph{Proceedings of The Web Conference 2020}, ser. WWW '20.\hskip 1em plus
  0.5em minus 0.4em\relax New York, NY, USA: Association for Computing
  Machinery, 2020, p. 1082–1092. [Online]. Available:
  \url{https://doi.org/10.1145/3366423.3380186}
\BIBentrySTDinterwordspacing

\bibitem{STGSA}
Z.~Wei, H.~Zhao, Z.~Li, X.~Bu, Y.~Chen, X.~Zhang, Y.~Lv, and F.-Y. Wang,
  ``Stgsa: A novel spatial-temporal graph synchronous aggregation model for
  traffic prediction,'' \emph{IEEE/CAA Journal of Automatica Sinica}, vol.~10,
  no.~1, pp. 226--238, 2023.

\bibitem{Zheng_Fan_Wang_Qi_2020}
\BIBentryALTinterwordspacing
C.~Zheng, X.~Fan, C.~Wang, and J.~Qi, ``Gman: A graph multi-attention network
  for traffic prediction,'' \emph{Proceedings of the AAAI Conference on
  Artificial Intelligence}, vol.~34, no.~01, pp. 1234--1241, Apr. 2020.
  [Online]. Available:
  \url{https://ojs.aaai.org/index.php/AAAI/article/view/5477}
\BIBentrySTDinterwordspacing

\bibitem{VAYGHAN2021110924}
\BIBentryALTinterwordspacing
L.~A. Vayghan, M.~A. Saied, M.~Toeroe, and F.~Khendek, ``A kubernetes
  controller for managing the availability of elastic microservice based
  stateful applications,'' \emph{Journal of Systems and Software}, vol. 175, p.
  110924, 2021. [Online]. Available:
  \url{https://www.sciencedirect.com/science/article/pii/S0164121221000212}
\BIBentrySTDinterwordspacing

\bibitem{SRIRAMA2020102629}
\BIBentryALTinterwordspacing
S.~N. Srirama, M.~Adhikari, and S.~Paul, ``Application deployment using
  containers with auto-scaling for microservices in cloud environment,''
  \emph{Journal of Network and Computer Applications}, vol. 160, p. 102629,
  2020. [Online]. Available:
  \url{https://www.sciencedirect.com/science/article/pii/S108480452030103X}
\BIBentrySTDinterwordspacing

\bibitem{YAN2021107216}
\BIBentryALTinterwordspacing
M.~Yan, X.~Liang, Z.~Lu, J.~Wu, and W.~Zhang, ``Hansel: Adaptive horizontal
  scaling of microservices using bi-lstm,'' \emph{Applied Soft Computing}, vol.
  105, p. 107216, 2021. [Online]. Available:
  \url{https://www.sciencedirect.com/science/article/pii/S1568494621001393}
\BIBentrySTDinterwordspacing

\bibitem{Wang2021}
S.~Wang, Z.~Ding, and C.~Jiang, ``Elastic scheduling for microservice
  applications in clouds,'' \emph{IEEE Transactions on Parallel and Distributed
  Systems}, vol.~32, no.~1, pp. 98--115, 2021.

\bibitem{fourati2022epma}
M.~H. Fourati, S.~Marzouk, and M.~Jmaiel, ``Epma: Elastic platform for
  microservices-based applications: Towards optimal resource elasticity,''
  \emph{Journal of Grid Computing}, vol.~20, no.~1, p.~6, 2022.

\bibitem{mekki2022microservices}
M.~Mekki, N.~Toumi, and A.~Ksentini, ``Microservices configurations and the
  impact on the performance in cloud native environments,'' in \emph{2022 IEEE
  47th Conference on Local Computer Networks (LCN)}, 2022, pp. 239--244.

\bibitem{fodor2022lsso}
B.~Fodor, L.~Toka, and B.~Sonkoly, ``Lsso: Long short-term scaling optimizer,''
  in \emph{2022 IEEE Conference on Network Function Virtualization and Software
  Defined Networks (NFV-SDN)}, 2022, pp. 52--58.

\bibitem{xu2022coscal}
M.~Xu, C.~Song, S.~Ilager, S.~S. Gill, J.~Zhao, K.~Ye, and C.~Xu, ``Coscal:
  Multifaceted scaling of microservices with reinforcement learning,''
  \emph{IEEE Transactions on Network and Service Management}, vol.~19, no.~4,
  pp. 3995--4009, 2022.

\bibitem{Chen2023}
X.~Chen, L.~Yang, Z.~Chen, G.~Min, X.~Zheng, and C.~Rong, ``Resource allocation
  with workload-time windows for cloud-based software services: A deep
  reinforcement learning approach,'' \emph{IEEE Transactions on Cloud
  Computing}, vol.~11, no.~02, pp. 1871--1885, apr 2023.

\bibitem{karypiadis2022scal}
E.~Karypiadis, A.~Nikolakopoulos, A.~Marinakis, V.~Moulos, and T.~Varvarigou,
  ``Scal-e: An auto scaling agent for optimum big data load balancing in
  kubernetes environments,'' in \emph{2022 International Conference on
  Computer, Information and Telecommunication Systems (CITS)}, 2022, pp. 1--5.

\bibitem{Rossi2020}
F.~Rossi, V.~Cardellini, and F.~L. Presti, ``Hierarchical scaling of
  microservices in kubernetes,'' in \emph{2020 IEEE International Conference on
  Autonomic Computing and Self-Organizing Systems (ACSOS)}, 2020, pp. 28--37.

\bibitem{Khaleq2021}
A.~A. Khaleq and I.~Ra, ``Intelligent autoscaling of microservices in the cloud
  for real-time applications,'' \emph{IEEE Access}, vol.~9, pp.
  35\,464--35\,476, 2021.

\bibitem{song2022automatic}
Y.~Song, C.~Li, K.~Zhuang, T.~Ma, and T.~Wo, ``An automatic scaling system for
  online application with microservices architecture,'' in \emph{2022 IEEE
  International Conference on Joint Cloud Computing (JCC)}, 2022, pp. 73--78.

\bibitem{baarzi2021showar}
\BIBentryALTinterwordspacing
A.~F. Baarzi and G.~Kesidis, ``Showar: Right-sizing and efficient scheduling of
  microservices,'' in \emph{Proceedings of the ACM Symposium on Cloud
  Computing}, ser. SoCC '21.\hskip 1em plus 0.5em minus 0.4em\relax New York,
  NY, USA: Association for Computing Machinery, 2021, p. 427–441. [Online].
  Available: \url{https://doi.org/10.1145/3472883.3486999}
\BIBentrySTDinterwordspacing

\bibitem{li2021joint}
Y.~Li, H.~Zhang, W.~Tian, and H.~Ma, ``Joint optimization of auto-scaling and
  adaptive service placement in edge computing,'' in \emph{2021 IEEE 27th
  International Conference on Parallel and Distributed Systems (ICPADS)}, 2021,
  pp. 923--930.

\bibitem{zhang2021sinan}
\BIBentryALTinterwordspacing
Y.~Zhang, W.~Hua, Z.~Zhou, G.~E. Suh, and C.~Delimitrou, ``Sinan: Ml-based and
  qos-aware resource management for cloud microservices,'' in \emph{Proceedings
  of the 26th ACM International Conference on Architectural Support for
  Programming Languages and Operating Systems}, ser. ASPLOS '21.\hskip 1em plus
  0.5em minus 0.4em\relax New York, NY, USA: Association for Computing
  Machinery, 2021, p. 167–181. [Online]. Available:
  \url{https://doi.org/10.1145/3445814.3446693}
\BIBentrySTDinterwordspacing

\bibitem{li2023topology}
X.~Li, J.~Zhou, X.~Wei, D.~Li, Z.~Qian, J.~Wu, X.~Qin, and S.~Lu,
  ``Topology-aware scheduling framework for microservice applications in
  cloud,'' \emph{IEEE Transactions on Parallel and Distributed Systems}, pp.
  1--17, 2023.

\end{thebibliography}

\end{document}